\documentclass{tlp}

\usepackage{latexsym}
\usepackage{amssymb}
\usepackage{pslatex}

\title[Representation Sharing for Prolog]{Representation Sharing for Prolog}

\author[Phuong-Lan Nguyen and Bart Demoen]
{
PHUONG-LAN NGUYEN\\
Institut de Math\'ematiques Appliqu\'ees, UCO, Angers, France \\
nguyen@ima.uco.fr
\and 
BART DEMOEN\\
Department of Computer Science, K.U.Leuven, Belgium\\
bart.demoen@cs.kuleuven.be
}

\usepackage{epsfig}
\usepackage{wrapfig}
\usepackage{fancyvrb}
\usepackage{subfigure}
\usepackage{latexsym}
\usepackage{pslatex}
\usepackage{amssymb}

\submitted{8 June 2010}
\revised{19 January 2011}
\accepted{30 May 2011}

\begin{document}

\date{}
\maketitle


\begin{abstract}
Representation sharing can reduce the memory footprint of a program by
sharing one representation between duplicate terms. The most
common implementation of representation sharing in functional
programming systems is known as hash-consing. In the context of
Prolog, representation sharing has been given little attention. Some
current techniques that deal with representation sharing are
reviewed. The new contributions are: (1) an easy implementation of
{\em input sharing} for {\em findall/3}; (2) a description of a {\em sharer}
module that introduces representation sharing at runtime. Their
realization is shown in the context of the WAM as implemented by
hProlog. Both can be adapted to any WAM-like Prolog
implementation. The sharer works independently of the garbage
collector, but it can be made to cooperate with the garbage
collector. Benchmark results show that the sharer has a cost
comparable to the heap garbage collector, that its effectiveness is
highly application dependent, and that its policy must be tuned to the
collector.
\end{abstract}





\section{Introduction}\label{intro}

Data structures with the same value during the rest of their common
life can share the same representation. This is exploited in various
contexts, e.g., by the {\em intern} method for Strings in Java, by
hash-consing in functional languages \cite{goto74}, and by
data deduplication during backup. In programming language
implementation, hash-consing is probably the best known representation
sharing technique: hash-consing was invented by Ershov in
\cite{ershov58} and used by Goto in \cite{goto74} in an implementation
of Lisp. Originally, hash-consing was performed during all term
creations so that no duplicate terms occurred during the execution of
a program. \cite{appelhashconsinggc} explores the idea of using
hash-consing only during generational garbage collection: the new
generation contains non-hash-consed terms, and on promotion to the
older generation, they are hash-consed: for the first time, a
representation sharing technique is cooperating with the garbage
collector. Our approach is most closely related to
\cite{appelhashconsinggc}, but also has some important differences.

\begin{sloppypar}
Representation sharing has been given little explicit attention in the
context of Prolog implementation. However, the issue pops up from time
to time. Here are some historical highlights:
\begin{itemize}
\item
in 1989, in his Diplomarbeit, Ulrich Neumerkel
\cite{neumerkeldiplomarbeit} mentioned how by applying DFA-minimization
to Prolog terms, certain programs can run in linear space (instead of
quadratic); there was no implementation; in Section \ref{benchmarks}, his
example program is used as a benchmark
\item
1991: \cite{VariableShunting} ends with the sentence: {\em It still
remains to be seen, however, what we meant by ``folding identical
structures''}; the current paper offers a solution to this mysterious
sentence
\item
in a 1995 comp.lang.prolog post, Edmund Grimley-Evans
\cite{findall1archive} asked for more sharing in {\em findall/3}, i.e., he
wanted the solution list of a call to {\em findall/3} to share with
the generator; {\em input sharing} \cite{findallwithoutfindall} does
exactly that; Section \ref{findall} describes input sharing more
precisely and how it can be implemented efficiently
\item
in a 2001 Logic Programming Pearl \cite{OKeefePearl}, R. O'Keefe
mentioned a {\em findall/3} query that could benefit from representation
sharing in the answers; as for the previous bullet, Section
\ref{findall} contains the solution
\item
in 2002, \cite{DemoenICLP2002fresh} gave a fresh view on garbage
collection for Prolog; it detailed a number of desirable (optimal)
properties of a garbage collector, one of which is the introduction of
representation sharing (albeit naming it differently)
\item
in May 2009, Ulrich Neumerkel posted an excerpt of his Diplomarbeit in
comp.lang.prolog and urged implementations to provide for more
representation sharing, either during unification, or during garbage
collection; he used the term {\em factoring}; we prefer {\em
  representation sharing}; the current paper is the result of
exploring its implementation issues
\end{itemize}
\end{sloppypar}

The paper is organized as follows: Section \ref{repsharversushash}
starts with describing what we mean by representation sharing. Section
\ref{somesharinginProlog} lists a number of more or less popular forms
of representation sharing in Prolog. Section \ref{findall} describes
how to retain {\em input sharing} for {\em findall/3} and evaluates our
implementation on a number of benchmarks.

Section \ref{generalities} sets the scene for the focus of the rest of
the paper: general sharing for Prolog. Section \ref{intuition}
forms the intuition on such sharing, while Section \ref{concepts}
introduces the notion of {\em absorption}: it shows when individual
cells can share their representation and the approximation that works
for us. It then lifts representation sharing from individual cells to
compound terms and discusses some properties of our notion of
representation sharing.
Section \ref{implementation} explains our implementation of
representation sharing based on the earlier decisions.
Section \ref{sharerbenchmarks} discusses the benchmarks and the
experimental results.
Section \ref{variants} shows extensions of the basic implementation,
variations and related issues. Section \ref{related} discusses related
work, and we conclude in Section \ref{conclusion}.

We have used hProlog 3.1.* as the Prolog engine to experiment with,
but it is clear that everything can be ported to other WAM-like
systems as well: we make that more explicit later on. hProlog is a
descendant of dProlog as described in \cite{wamvariations}. SICStus
Prolog 4.1.1 serves as a yardstick to show that the hProlog time and
space figures are close to a reliable state of the art system.
All benchmarks were run on an Intel Core2 Duo Processor T8100 2.10
GHz.

We assume the reader to be familiar with the WAM \cite{wam:hassan,DHWa83}
and Prolog \cite{ClMe84}. We use the term {\em heap} when others use
{\em global stack}, i.e., the place where compound terms are
allocated. We use {\em local stack} and {\em environment stack}
interchangeably and denote it by LS in pictures.

\section{Representation Sharing versus Hash-Consing}\label{repsharversushash}

\begin{figure}[h]
\begin{centering}
{\epsfig{file=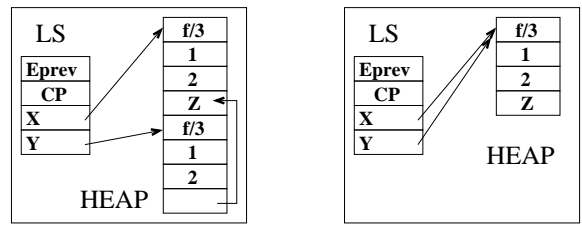,width=0.6\textwidth}}
\caption{Representation Sharing}
\label{fig1}
\end{centering}
\end{figure}

Consider the predicates main1 and main2 defined as
\begin{Verbatim}[fontsize=\small, frame=single,samepage=true]
main1 :-                          main2 :-
       X = f(1,2,Z),                     X = f(1,2,Z),
       Y = f(1,2,Z),                     X = Y,
       use(X,Y).                         use(X,Y).
\end{Verbatim}
In a naive\footnote{I.e., an implementation without compile time
  common subexpression elimination; however, note the danger of such
  optimization in the presence of destructive assignment: see Section
  \ref{mutable} }
implementation, the execution of ?- main1. just before the call to
{\em use/2}, results in a memory situation as in the left of Figure
\ref{fig1}. In this figure, the heap cell with Z is a self-reference
in the WAM.
Clearly, the terms X and Y are exactly the same ever after they have
been created, and therefore they can share the same representation:
that sharing can be seen in the right of Figure \ref{fig1} and in the
code for the predicate main2.

Hash-consing is usually associated with the technique that keeps a
hash table of terms and during term creation checks whether a term is
new or exists already in the hash table.

An implementation with hash-consing usually changes the representation
of terms, and consequently the code that deals explicitly with this
representation. For Prolog the affected code would be general
unification and built-in predicates. That is too intrusive for our
aims: we intend our implementation of representation sharing to be
easy to integrate in other Prolog systems and there should be no
global impact. So, we will keep the usual (WAM) term representation
and do not touch any part of the implementation, except for the {\em
  sharer} module that introduces representation sharing. Given the
complexity of current Prolog systems, this seems to us the only way to
make representation sharing accepted by implementors.

\subsection*{Some Forms of Representation Sharing for Prolog}\label{somesharinginProlog}

Prolog implementations already provide some specific representation
sharing. Here are a few examples:

\begin{sloppypar}
\begin{itemize}

\item
in older implementations, the predicate {\em copy\_term/2} copies
ground terms; in newer implementations ---starting probably with
SICStus Prolog \cite{matsphd}--- {\em copy\_term/2} avoids copying
ground (sub)terms; this means that the second argument can have some
representation sharing with the first argument; however, note that
{\em mutable} ground terms must be copied by {\em copy\_term/2},
because otherwise sharing would become observable at the program
level; we discuss this issue further in Section \ref{mutable}

\item
some programs contain ground terms at the source level; a typical
example is the second argument of a goal like {\em
  member(Assoc,[fx,fy,xfx,xfy,yfx,xf,yf])}; ECLiPSe
\cite{Wallace97eclipse} pre-allocates such ground terms, and makes
sure that any time such a fact or goal is called, the ground term is
re-used; Mercury performs this compile-time optimization as well

\item
when two terms are unified, they can share a common representation in
the forward execution; at various stages in its life, BinProlog
\cite{Tarau91:JAP} enforced such sharing by (in WAM speak) redirecting
the S-tagged pointer of one of the two terms and (conditionally)
trailing this change so that on backtracking it can be undone; if
trailing is not needed, then the savings can be huge; otherwise, the
locality of access can be improved, but memory and time savings can be
negative; a similar technique was already used for strings only in the
Logix implementation of Flat Concurrent Prolog \cite{logixFCP}

\end{itemize}
\end{sloppypar}

In each of the above cases, the implementor of the Prolog system
decided for more representation sharing than would be the case in
a more straightforward implementation.
Application programmers and library developers usually take care as
well to let their runtime data structures share common parts.

In the above, {\em copy\_term/2} and unification are built-in
predicates that have a chance to increase representation sharing. In
Section \ref{findall}, {\em findall/3} is added to this shortlist.

\section{Input Sharing for {\em findall/3}}\label{findall}

In \cite{findallwithoutfindall}, the notion of {\em input sharing} was
introduced in the context of {\em findall/3}.  Input sharing consists
of a solution in the output from {\em findall/3} (its third argument)
sharing with the input to {\em findall/3} (its second argument).

Later, in the Logic Programming Pearl \cite{OKeefePearl} it is
suggested that {\em findall/3} could avoid repeatedly copying the same
terms over and over again: this would improve the space complexity of
some queries that use {\em findall/3}, from $O(n^2)$ to
$O(n)$. However, R. O'Keefe suggests that hash-consing should be used,
with the consequence that the time complexity remains the same: our
implementation of {\em input sharing} ---which is exactly what is
needed here--- improves both the time and space complexity. The example
used in \cite{OKeefePearl} is rather complicated, so for now, we use
as an illustration a piece of simple Prolog code that was posted in
\cite{findall1archive}; we changed the names of the predicates and
variables.

\begin{Verbatim}[fontsize=\small, frame=single,samepage=true]
       findall_tails(L,Tails) :- findall(Tail,is_tail(L,Tail),Tails).
       
       is_tail(L,L).
       is_tail([_|R],L) :- is_tail(R,L).
       
       all_tails([],[[]]).
       all_tails(L,[L|S]) :- L = [_|R], all_tails(R,S).
\end{Verbatim}

Clearly, goals of the form 
?- findall\_tails(L,Tails).
and
?- all\_tails(L,Tails).
with a ground argument L succeed with the same answer Tails. E.g.,

\begin{Verbatim}[fontsize=\small, frame=single,samepage=true]
       ?- findall_tails([1,2,3],Tails).
       Tails = [[1,2,3],[2,3],[3],[]]
\end{Verbatim}


The usual implementation of {\em findall/3} copies over and over again parts of
the input list L, and this results in quadratic behavior (in the
length of L) for {\em findall\_tails/2}, while {\em all\_tails/2} is linear, both
in space and time! Clearly, with enough {\em input sharing} the
{\em findall\_tails/2} query could be linear.

In the following sections we show how a traditional {\em findall/3}
implementation in the context of the WAM can be easily adapted to
cater for input sharing. An alternative copy-once implementation of
{\em findall/3} is also shown.

Before going into the details, it is worth pointing out the
limitations of input sharing. Clearly, if L is a list with non-ground
elements, the two queries
\begin{Verbatim}[fontsize=\small, frame=single,samepage=true]
  ?- findall_tails(L,Tails).                ?- all_tails(L,Tails).
\end{Verbatim}
yield different answers. The first query makes fresh variants of the
variables in each of the solutions in Tails, while the second query
does not. As an example:

\begin{Verbatim}[fontsize=\small, frame=single,samepage=true]
?- findall_tails([X,Y,Z],L),          ?- all_tails([X,Y,Z],L),
   numbervars(L,0,_).                    numbervars(L,0,_).

                                      X = A   Y = B   Z = C
L = [[A,B,C],[D,E],[F],[]]            L = [[A,B,C],[B,C],[C],[]]
\end{Verbatim}

This means we can use an input sharing version of {\em findall/3} when the
arguments of the generator are either ground or free: the danger is
only in terms containing variables.

\subsection{The Implementation of {\em findall/3}}

The hProlog implementation of {\em findall/3} follows the same pattern as
in many systems:

\begin{Verbatim}[fontsize=\small, frame=single,samepage=true]
       findall(Template,Generator,SolList) :-
              findall_init(Handle),
              (
                call(Generator), 
                findall_add(Template,Handle),
                fail
              ;
                findall_get_solutions(SolList,Handle)
              ).
\end{Verbatim}

For simplicity, we have left out all error checking and error recovery
code. The predicate {\em findall\_init/1} returns a handle, so that the
particular invocation of {\em findall/3} is identified: this is used for
correct treatment of nested calls to {\em findall/3}. {\em findall\_add/2} uses
that handle, and copies the Template to a temporary
zone. {\em findall\_get\_solutions/2} uses the handle as well: it retrieves
the complete list of solutions from the temporary zone and unifies it
with the third argument to {\em findall/3}.

The next section describes how to turn this code into code that shares
the input.

\subsection{The basic Idea of Input Sharing for {\em findall/3}}

\begin{sloppypar}
The predicate {\em findall\_add/2} in our implementation of {\em findall/3} is just a
version of {\em copy\_term/2}: at the implementation level, they both use
the same C function for the actual copying. The same is true for
{\em findall\_get\_solutions/2}.
\end{sloppypar}

The first idea might be to use an implementation of {\em copy\_term/2} that
avoids copying ground terms. However, in the context of {\em findall/3},
groundness is not enough: the ground term must also be {\em old
  enough}, so that backtracking (over the Generator) cannot alter
it. To be more precise, anything ground that survives backtracking
over the Generator need not be copied by {\em findall\_add/2}. Or put still
another way: anything ground before the call to
findall(Template,Generator,SolList) need not be copied by
{\em findall\_add/2}.

Such terms can be recognized easily: their {\em root} resides in a
heap segment that is not younger than the call to {\em findall/3}.

So we need to be able to identify the older heap part relevant to a
particular call to {\em findall/3}. That is quite easy in the WAM: we just
remember the relevant heap pointer!

\subsection{{\em findall/3} with Input Sharing: the Implementation}\label{impl}

We use two new low-level built-in predicates:
\begin{sloppypar}
\begin{itemize}
\item {\bf current\_heap\_top(--)}: unifies the argument with (an
abstraction of) the current value of the heap pointer H
\item {\bf set\_copy\_heap\_barrier(+)}: sets a global
(C-)implementation variable (named {\em copy\_heapbarrier}) to the
heap pointer value corresponding to its argument
\end{itemize}
\end{sloppypar}

The following code shows how the new built-ins are used:

\begin{Verbatim}[fontsize=\small, frame=single,samepage=true]
       sharing_findall(Template,Generator,SolList) :-
              current_heap_top(Barrier),
              findall_init(Handle),
              (
                call(Generator), 
                set_copy_heap_barrier(Barrier),
                findall_add(Template,Handle),
                fail
              ;
                set_copy_heap_barrier(Barrier),
                findall_get_solutions(SolList,Handle)
              ).
\end{Verbatim}

An additional small change needs to be made to the implementation
of {\em findall\_add/2} (and {\em findall\_get\_solutions/2}) as well: when a term
is about to be copied and it is older than {\em copy\_heapbarrier},
only the root pointer to this term is copied. It amounts to adding a
statement like
\begin{Verbatim}[fontsize=\small, frame=single,samepage=true]
   if (struct_addr < copy_heapbarrier) { *whereto = struct_addr; continue; }
\end{Verbatim}
at a few places in the C code of {\em copy\_term/2}: this piece of code
just copies the top pointer of the structured term instead of copying
it recursively. The C variable {\em struct\_addr} holds the address
of the structure about to be copied.

For explanatory reasons, we have shown the implementation of {\em
  sharing\_findall/3} as a variant of the basic implementation of {\em
  findall/3} using two new built-ins. However, one can also fold the
functionality of these new built-ins into adapted versions of
findall\_[init, add, get\_solutions]: the top-of-heap at the moment of
calling {\em sharing\_findall/3} is then stored in the data structures
belonging to that particular call. This top-of-heap at the moment of
calling {\em sharing\_findall/3} must also be appropriately treated by
the garbage collector.

\subsection{An Example}

\begin{figure}[h]
\begin{centering}
{\epsfig{file=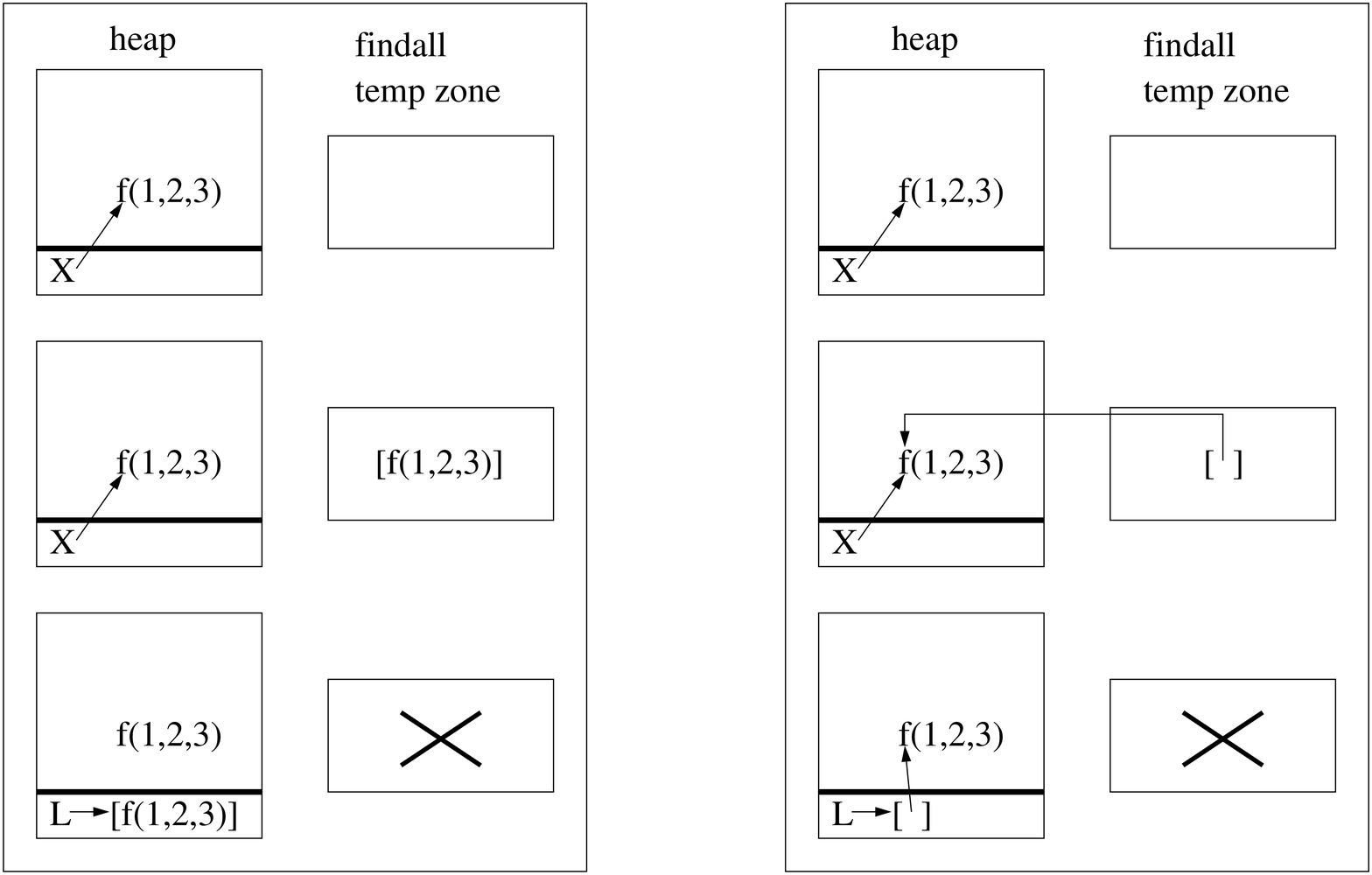,width=1.0\textwidth}}
\caption{Findall without and with Input Sharing}
\label{findallfig1}
\end{centering}
\end{figure}

The heap and temporary findall zone are shown in Figure \ref{findallfig1} for
the very simple query

\begin{Verbatim}[fontsize=\small, frame=single,samepage=true]
       ?- findall(X, X=f(1,2,3), L).
\end{Verbatim}

The left part of the picture shows three snapshots during the execution
of the query without input sharing. The right part shows the
corresponding snapshots with input sharing.

The snapshots are taken 
\begin{itemize}
\item just before {\em findall\_add/2} is executed: the temporary zone is
  still empty
\item just after {\em findall\_add/2} is executed; at the left, the temporary
zone contains a copy of the term f(1,2,3); at the right, there is a
pointer to the term on the heap
\item just after {\em findall\_get\_solutions/2} is executed: the temporary
  zone can be discarded; at the left, the solution list contains a
copy of f(1,2,3); at the right, there is just a pointer to the old
term on the heap
\end{itemize}

The space savings are clear.

\subsection{Copy-once {\em findall/3}}

The usual implementation of {\em findall/3} copies the solutions twice.
BinProlog was probably the first implementation copying the solution
only once, by means of a technique named {\em heap lifting} or more
popularly a {\em bubble in the heap}
\cite{ecologicalPaul@IWMM-92}. Currently, the BinProlog
implementation \cite{padl09inter} relies on {\em engines} for
{\em findall/3}. Mercury also uses a copy-once findall (named {\em solutions/2}):
as Mercury relies on the Boehm-collector \cite{hansboehm}, there is no memory
management hassle with a bubble in the heap.

It is rather easy to implement a copy-once {\em findall/3} in any
Prolog system that has non-backtrackable destructive assignment (with
{\em nb\_setarg/3}) as in hProlog or SWI-Prolog
\cite{swiprolog}\footnote{Note that SWI-Prolog uses {\em nb\_linkarg/3}
  as the name for hProlog's {\em nb\_setarg/3}}:

\begin{Verbatim}[fontsize=\small, frame=single,samepage=true]
       copy_once_findall(Template,Generator,SolList) :-
              Term = container([]),
              (
                call(Generator),
                Term = container(PartialSolList),
                copy_term(Template,Y),
                nb_setarg(1,Term,[Y|PartialSolList]),
                fail
              ;
                Term = container(FinalSolList),
                reverse(FinalSolList,SolList)
              ).
\end{Verbatim}

As before, this code can be enhanced with the newly introduced
built-ins to yield a copy\_once\_sharing\_findall. If copying the
solutions dominates the execution, the copy-once {\em findall/3} is about
twice as fast as the regular {\em findall/3}. However, its main drawback is
that it consumes (for the benchmarks below) about three times as much
heap space. The reason is that {\em nb\_setarg/3} must freeze the heap
if its third argument is a compound term. The heap-lifting technique
(which we have not implemented in hProlog) does not have this
drawback.

\subsection{Experimental Evaluation}

Input sharing improves (sometimes) the complexity (space and time),
and the constant overhead is really very small, as can be judged from
the changes needed to implement it. One could therefore argue that
benchmarks are not needed. Even so, we present two benchmarks: one is
the {\em findall\_tails/2} example (see Section \ref{tailsbenchmark}). We
start with a {\em findall/3} related query from \cite{OKeefePearl}: this
pearl is about tree construction and traversal. It contains the
following text:

\begin{itemize}
\item[]
{\em Query q1\footnote{f1(N) in the Appendix} requires at least
  $O(n^2)$ space to hold the result.}  {\em If findall/3 copied terms
  using some kind of hash consing, the space cost could be reduced to
  $O(n)$, but not the time cost, because it would still be necessary
  to test whether a newly generated solution could share structure
  with existing ones.}
\end{itemize}

Note that the $n$ above is the number of nodes in the tree, not the
tree depth: the number of nodes is roughly $4^{depth}$ where {\em
depth} is the depth of the tree.

We needed to make a slight change to the program from
\cite{OKeefePearl}: in its original form it contains a {\em mk\_tree/2}
predicate defined as

\begin{Verbatim}[fontsize=\small, frame=single,samepage=true]
       mk_tree(D, node(D,C)) :-
              (   D > 0 ->
                  D1 is D - 1,
                  C = [T,T,T,T],
                  mk_tree(D1, T)
              ;   C = []
              ).
\end{Verbatim}
Because of the conjunction {\em C = [T,T,T,T], mk\_tree(D1, T)}, the
heap representation of the constructed tree is linear in the first
argument D, even though it has an exponential number of nodes: indeed,
the constructed tree has a lot of internal sharing. Such internal
sharing is retained by most reasonable implementations of {\em copy\_term/2}
and by {\em findall\_add/2}.\footnote{A notable exception is Yap.}

In order to test what \cite{OKeefePearl} really meant, we have changed
the particular conjunction to
\begin{Verbatim}[fontsize=\small, frame=single,samepage=true]
         C = [T1,T2,T3,T4],
         mk_tree(D1, T1), mk_tree(D1, T2),
         mk_tree(D1, T3), mk_tree(D1, T4)
\end{Verbatim}
so that the size of the representation of the tree is linear in the
number of nodes in the tree (and exponential in D). This code rewrite
achieves the desired effect because Prolog systems typically don't
perform the analysis needed to notice that T1, T2, T3 and T4 are
declaratively the same value, and neither is this detected at
runtime. See the Appendix for all code necessary to run the benchmark.

\subsubsection{The modified Tree Benchmark: Results}

Table \ref{treecombined} shows timings (when considered meaningful)
and space consumption for queries {\em ?- f1(Depth)} with different
values of Depth. Times are reported in milliseconds, space in
bytes. We have chosen SICStus Prolog for comparison with another
system because the SICStus Prolog implementation performed better and
more reliably than the other systems we tried. Moreover, the trend of
the measurements with other systems was basically the same.

The timings without sharing do not show anything interesting
complexity-wise: neither of the implementations without input sharing
can deal with more than about 5000 nodes. The input sharing
implementation on the other hand can go easily up to one million
nodes. The heap consumption columns give a good picture of how the
heap size grows: the non-input-sharing implementations show a
quadratic dependency of the heap consumption on the number of
nodes. Only {\em hProlog input sharing} shows a linear
dependency.

\begin{table}[ht]
\begin{center}
\begin{tabular}{|r||r|r||r|r||r|r|}
\hline
      & \multicolumn{2}{c||}{hProlog} & \multicolumn{2}{c||}{hProlog input sharing} & \multicolumn{2}{c|}{SICStus Prolog}\\
\cline{2-7}
Depth & time & space & time & space & time & space \\ 
\hline
1        &         &  964              &               & 368           &            &980       \\
2        &         &    11332          &               & 1840          &            &10964     \\
3        &         &    158020         &               & 9776          &            &155348    \\
4        &         &    2394436        &               & 49712         &            &2379476   \\
5        &    156  &    37599556       &               & 242224        &  250       &37523156  \\
6        &    2616 &    598030660      &               & 1143344       &  6820      &597659348 \\
7        &         &                   &               & 5272112       &            &          \\
8        &         &                   &  156          & 23884336      &            &          \\
9        &         &                   &  652          &106721840      &            &          \\
10       &         &                   &  2804         &471626288      &            &          \\
\hline
\end{tabular}
\caption{Heap consumption in bytes and time in msecs  for the tree
  benchmark}\label{treecombined}
\end{center}
\end{table}

Table \ref{treecombined} shows clearly that our simple implementation
to enforce input sharing is very effective and performs actually
better than hoped for in \cite{OKeefePearl}. Indeed, we achieve linear
space {\bf and} time complexity for the f1(Depth) query. Hash-consing
would not be able to do that.

\subsection{The {\em tails} Benchmark}\label{tailsbenchmark}

Table \ref{tailscombined} shows the space consumption for the tails
benchmark. The timings are meaninglessly small for the variants with
sharing, and therefore only shown for the regular findall columns. The
{\em Length/1000} column indicates the length of the ground input list
L to queries of the form {\em ?- all\_tails(L,Tails)} and {\em ?-
  [sharing\_]findall(Tail,is\_tail(L,Tail),Tails)}.

\begin{table}[ht]
\begin{center}
\begin{tabular}{|r|r|r|r|r||r|r|r|}
\hline
       & \multicolumn{4}{c||}{hProlog} & \multicolumn{3}{c|}{SICStus Prolog}\\ \hline
       & \multicolumn{2}{c|}{regular} & \multicolumn{1}{c|}{findall with} & \multicolumn{1}{c||}{all\_tails}& \multicolumn{2}{c|}{regular       } & \multicolumn{1}{c|}{all\_tails}\\
Length & \multicolumn{2}{c|}{findall} & \multicolumn{1}{c|}{input sharing} & \multicolumn{1}{c||}{}           & \multicolumn{2}{c|}{findall       } & \multicolumn{1}{c|}{              }\\
\cline{2-8}
/1000 & time & space & space & space & time & space & space \\ 
\hline
1        &112    &  26034     &  8   &  8   &   120  &  26034  &  8   \\
2        &444    & 104068     & 16   & 16   &   490  & 104068  & 16   \\
3        &1028   & 234102     & 24   & 24   & 1160   & 234102  & 24   \\
4        &1920   & 416136     & 32   & 32   & 2130   & 416136  & 32   \\
5        &       &            & 40   & 40   & 3650   & 650170  & 40   \\
6        &       &            & 48   & 48   & 5080   & 936204  & 48   \\
7        &       &            & 56   & 56   &        &         & 56   \\
8        &       &            & 64   & 64   &        &         & 64   \\
9        &       &            & 72   & 72   &        &         & 72   \\
10       &       &            & 80   & 80   &        &         & 80   \\
100      &       &            &800   &800   &        &         &800   \\
1000     &       &            &8000  &8000  &        &         &8000  \\
\hline
\end{tabular}
\caption{Heap consumption in KiB and time in msecs - tails
  benchmark}\label{tailscombined}
\end{center}
\end{table}

{\em findall/3} with input sharing clearly beats the {\em findall/3}
without input sharing. SICStus Prolog can do larger sizes with the
ordinary {\em findall/3} implementation than hProlog: the latter runs
out of memory earlier because of its different memory allocation and
heap garbage collection policy.

We have tried to measure the overhead of our method, but it is too
small to show up meaningfully in any of our experiments.

\subsection{Conclusion on Input Sharing for {\em findall/3}}

Already in \cite{findall1archive} there was a demand for sharing
between the input to {\em findall/3} and its output. Also \cite{OKeefePearl}
points out that this would be beneficial to some programs.
Optimal input sharing would attempt to share all (sub)terms that are
ground just before the call to {\em findall/3}. Checking this at
runtime can be involved and costly. Our implementation approximates
that by just checking that the root of a term is old enough,
and relying on the programmer (or some other means) to use {\em
  sharing\_findall/3} only when this simple check implies that the
whole term was ground at the moment of the call to {\em
  sharing\_findall/3}. This is in particular true in the common case
that the generator of {\em findall/3} (its second argument) is a
goal of which every argument is ground or free: that was the case for
our benchmark {\em findall\_tails/2}. That condition on the generator
can be easily checked before calling {\em sharing\_findall/3} and
could also be derived by program analysis.

\begin{sloppypar}
Our approach does not implement {\em solution sharing}: hash-consing,
or maybe even better tries, could do the job. Sections
\ref{generalities} and later provide a more general and lightweight
solution to representation sharing.
\end{sloppypar}

In \cite{OKeefePearl}, one can also read:
\begin{itemize}
\item[]
{\em One referee suggested that Mercury's `solutions/2' would be
cleverer. A test in the 0.10 release showed that it is not yet clever
enough.}
\end{itemize}

As Mercury \cite{zoltan:mercury} relies on the Boehm-allocator and
-collector for its memory management, it is quite difficult to devise
a simple dynamic test whether a (ground) term is old enough: on the
whole, a Mercury implementation does indeed not benefit from keeping
the address order of terms consistent with their age. On the other
hand, in the WAM, such a test comes natural with the needs of a strict
heap allocation discipline and conditional trailing.

As a conclusion, we think we have succeeded in providing input sharing
for {\em findall/3} with minimal change to the underlying Prolog execution
engine: any Prolog implementation with a heap allocation strategy
similar to the WAM can incorporate it easily. How to present the
functionality in a safe way to the user is a language design issue and
as such beyond the scope of this paper.

\section{General Representation Sharing for Prolog}\label{generalities}

\cite{appelhashconsinggc} adapts a copying collector to perform
hash-consing for the data in the older generation. Since we would like
our implementation of representation sharer to be a model for other
Prolog implementations, we cannot just copy that idea. Indeed,
hash-consing requires a serious adaptation of the term
representation, and moreover Prolog systems typically have sliding
collectors, the exceptions being hProlog and BinProlog. Therefore we
want to investigate representation sharing in a way that does not
require a change in term representation, and that is independent of
the details of the garbage collector: this will make it easier for
Prolog systems to implement their own sharing module based on our
experience. \cite{appelhashconsinggc} argues that garbage collection
time is a good moment to perform hash-consing, but there is no
inherent need to do it only then. Still, we agree basically with
\cite{appelhashconsinggc}: it is better to avoid putting any effort in
sharing with dead terms.

We use as Prolog goals in the examples {\em share} and {\em gc}: the
former performs representation sharing, the latter just performs
garbage collection. By keeping the two separated, the issues become
clearer, i.e., we make no assumptions on the workings of the garbage
collector.

\cite{bakerwarpspeed} shows that the combination of tabling and
hash-consing is particularly powerful: since duplicate terms do not
occur, equality of terms can be decided by a single pointer comparison
instead of by traversing the whole terms. However, in that context and
in its original form, hash-consing guarantees representation sharing
all the time, while that is not our aim. Unfortunately,
\cite{bakerwarpspeed} does not show experimental data for hash-consing
without tabling.

\section{Representation Sharing in Prolog: Examples}\label{intuition}

Two issues make representation sharing in Prolog-like languages
different from other languages: the logical variable and backtracking.
Subsequent subsections show by example how these affect the
possibilities for representation sharing.

\subsection{Sharing within the same Segment}\label{samesegment}

The first example in Section \ref{repsharversushash} shows the 
simplest case of sharing: the two terms are identical, in the same heap
segment (as delimited by the HB pointers in the choicepoints) and
ground at creation time.

The next example shows that identical ground terms in the same segment
cannot always share their representation:
\begin{Verbatim}[fontsize=\small, frame=single,samepage=true]
       main3 :-
              T1 = f(a),
              T2 = f(X),
              (
               X = a, share
              ;
               write(T1 \== T2)
              ).
\end{Verbatim}
While executing the query ?- main3, just before the execution of {\em
share}, the terms T1 and T2 are identical, ground, and they are
completely within the same segment. However, it would be wrong to make
them share their representation, since in the failure continuation,
they are no longer identical. Loosely speaking, the occurrence of
trailed variables in a term makes the term unsuitable for representation
sharing.

\subsection{Sharing between Segments}

The previous examples dealt with representation sharing of terms that
live in the same segment. The next example shows an issue with
representation sharing of terms that live in different segments. Since
we do not want to mix this issue with trailed heap locations, the
example works with ground terms.

\begin{Verbatim}[fontsize=\small, frame=single,samepage=true]
       main4 :-
              T1 = f(a),
              (
               T2 = f(a), share, use(T1,T2)
              ;
               use(T1)
              ).
\end{Verbatim}

\begin{figure}[h]
\begin{centering}
{\epsfig{file=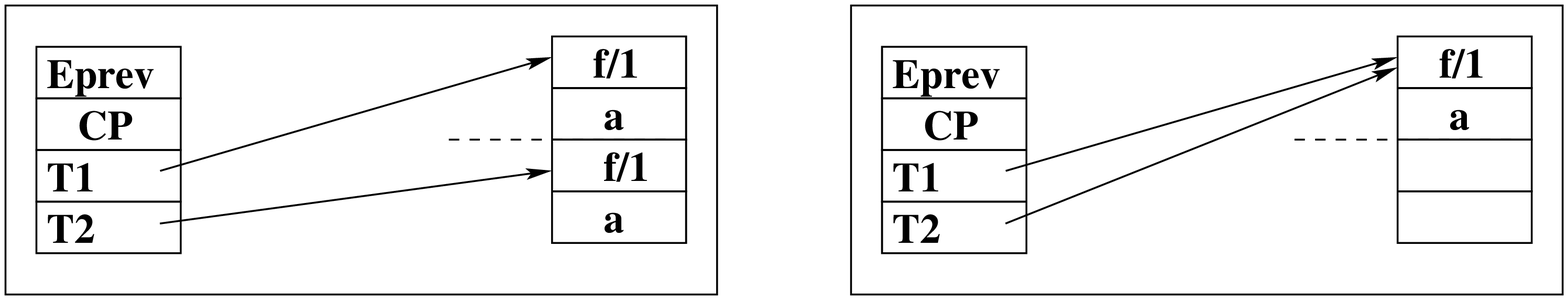,width=0.9\textwidth}}
\caption{Representation Sharing of two Terms in different Segments}
\label{fig2}
\end{centering}
\end{figure}

During the execution of the query ?- main4, T1 and T2 live in two
different segments. T1 lives in the oldest segment, as seen in the
left of Figure \ref{fig2}\footnote{The dashed line indicates the heap
segment barrier}. Since T1 is used after backtracking, the natural
thing is to keep the representation of the oldest term, because it
potentially lives longest. So the introduced sharing representation is
as in the right of Figure \ref{fig2}.
Alternatively, one could use as shared representation the one in the
younger segment, but then the heap should be frozen, so that on
backtracking the value of T1 does not get lost. We consider this a bad
alternative, but a slight variation on the same example shows that the
choice is not so clear cut:
\begin{Verbatim}[fontsize=\small, frame=single,samepage=true]
main5 :-                                    main6 :-
       T1 = f(a),                                  T1 = f(a),
       use(T1),                                    use(T1),
       (                                           (
        T2 = f(a), share, gc, use(T2)               T2 = f(a), gc, use(T2)
       ;                                           ;
        dontuseT1                                   dontuseT1
       ).                                          ).
\end{Verbatim}
The code of main5 and main6 differs only in the call to {\em share} in
main5.

\begin{itemize}
\item {\bf with sharing in main5}: share keeps one representation of
f(a) and puts it in the oldest segment; gc cannot reclaim that
representation, because T2 is not dead; after backtracking to
dontuseT1, the f(a) term is still on the heap
\item {\bf without sharing in main6}: at the point gc kicks in, T1 is
unreachable and its representation disappears; this means that after
backtracking to dontuseT1, the heap is empty
\end{itemize}

This example shows that representation sharing between terms in
different segments can lead to a higher heap consumption, or more
invocations of the garbage collection.

Finally, it is clear that mutable terms should not share their
representation: it is in general impossible to know whether two
mutable terms will be identical for the rest of their common
lifetime. We deal with mutable terms in more detail in Section
\ref{mutable}.

\section{Sharable Terms and Absorption}\label{concepts}

The examples in the previous section give some intuition on what we
mean by representation sharing, and also about its pitfalls. The
examples also have indicated that we are working towards an
implementation of a sharer that introduces sharing between two terms
T1 and T2 by keeping the representation of one of them, say T1, and
making T2 point to it. We coin this process {\em T1 absorbs T2}. This
leads naturally to considering the notion T1 {\em can absorb} T2.

The most general definition of {\em T1 can absorb T2} would be that
the sequence of solutions to the running program does not change by
letting T1 absorb T2. That condition is of course not decidable, so we
need a workable approximation to it.

The next sections explore the notion {\em can absorb} further, first by
focussing on representation sharing for individual heap cells and then
by considering compound terms.

\subsection{Representation Sharing for Individual Heap Cells}\label{individualcell}

It pays off to study the most basic representation sharing of all:
between two individual heap cells.

Clearly, when two cells, say c1 and c2, have different contents (and
are live), neither of them can absorb the other. And when the two
cells have identical addresses, they have absorbed each other
already. So, we are left with the possibilities that
\begin{itemize}
\item c1 and c2 are in the same heap segment or not
\item c1 and/or c2 is trailed or not
\end{itemize}
Without loss of generality, we assume that c1 is older than c2.

\begin{figure}[h]
\begin{centering}
{\epsfig{file=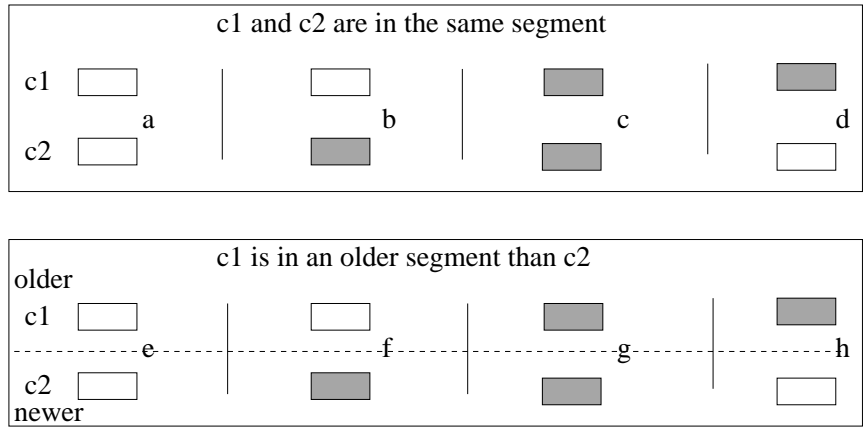,width=0.9\textwidth}}
\caption{The 8 combinations of two cells}
\label{fig3}
\end{centering}
\end{figure}

This results in the eight combinations shown in Figure \ref{fig3}: a
trailed cell is shaded. The contents of the two cells at the moment of
the snapshot is the same, but shaded cells will be set to {\em free}
(a self-reference in the WAM) on backtracking to the appropriate
choicepoint. The horizontal dashed lines now indicate one or more heap
segment separations. The vertical lines just separate the different
cases.
\begin{enumerate}
\item[{\bf a}:]
c1 can absorb c2 and also vice versa, because the two cells have an
identical contents, and that will remain so in the forward and in the
backward computation
\item[{\bf bcd}:]
in the forward computation, the two cells remain identical, but not
after backtracking; so no representation sharing can take place, and
neither can absorb the other
\item[{\bf e}:]
on backtracking, c2 {\em dies} before the older cell c1, but for the
duration of their common life, the two cells are identical, so
representation sharing is allowed: c1 can absorb c2, but not the other
way around
\item[{\bf fg}:]
these cases are similar to cases {bf bc} above: as soon as one of the
trailed cells is untrailed by backtracking, the contents of c1 and c2
differ; therefore representation sharing is not allowed; neither can
absorb the other
\item[{\bf h}:]
there are two possibilities now:
\begin{enumerate}
\item
at the moment the older cell is untrailed, backtracking also recovers
the segment in which the newer cell resides; this means that the newer
cell dies, so the fact that the older cell is set to {\em free} does
not prevent representation sharing; so c1 can absorb c2 (and not the
other way around); this happens if c1 was trailed before the segment
of c2 is final, or to put it differently: if the moment of trailing c1
is not after the segment of c2 is closed by a choicepoint; i.e., if c2
dies not later than c1 is untrailed, c1 can absorb c2
\item
otherwise, representation sharing is disallowed; neither cell can
absorb the other
\end{enumerate}

\end{enumerate}


Anticipating an implementation, we notice that it is important to be
able to check quickly whether a cell is trailed. One bit
---appropriately placed--- is enough for that: that bit could be in
the heap cells themselves, or it could be allocated in an array
parallel to the heap. This would make cases {\bf a} and {\bf e} easy
to identify.

To detect case {\bf h(a)} however, we also need to retrieve quickly
from a heap address, the heap segment number in which it was
trailed. That requires more setup, and it would slow down the
sharer. We think the expected gain in space too small to make this
worthwhile. Instead, we went for disallowing sharing in case {\bf
  h(a)}, so that our notion of {\em can absorb} becomes quite
simple and leads to a simple decision procedure. In the following
piece of code, pc1 and pc2 are pointers to heap cells c1 and c2:

\begin{Verbatim}[fontsize=\small, frame=single,samepage=true]
       boolean can_absorb(cell *pc1, cell *pc2)
       {
         if (*pc1 != *pc2) return(FALSE);
         if (trailed(pc1)) return(FALSE);
         if (trailed(pc2)) return(FALSE);
         return(pc1 < pc2);
       }
\end{Verbatim}

If cell c1 can absorb cell c2, every (tagged) pointer to c2 can be
changed into a (tagged) pointer to c1: this change does not affect the
outcome of the execution. Note that it is immaterial whether the cell
containing the (tagged) pointer to c2 is trailed or not.

Since trailing prevents a cell from being able to absorb, or being
absorbed, it is in the interest of maximizing the chances for
representation sharing to keep the trail {\em tidy}: this is in many
Prolog systems done at the moment a cut ({\em !/0}) is executed. Also during
garbage collection, the trail can be tidied.

\subsection{Representation Sharing for Compound Terms}\label{compoundterms}

The representation of a compound term with principal functor {\em
  foo/n} in the WAM is an S-tagged pointer to an array of (n+1)
contiguous heap cells, the first of which contains {\em foo/n}, and the next
n cells contain one cell of the representation of one argument
each. We name this array of (n+1) heap cells the {\em body} of the
term.

\begin{figure}[h]
\begin{centering}
{\epsfig{file=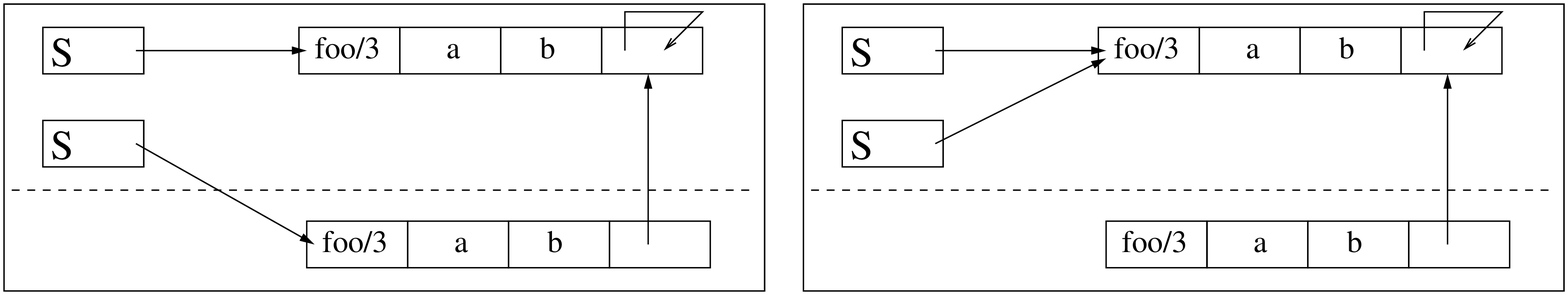,width=1.0\textwidth}}
\caption{Left: the bodies fulfill the conditions for representation sharing.
~~~~~~~~~~~~~~~~~~~~~~~~~~~~~~~~ $~~~~~~~~~~~~~$Right: sharing has been performed}
\label{fig6}
\end{centering}
\end{figure}

The idea of one term absorbing the other is that after absorption,
there is only one body instead of two, but there are still two cells
with an S-tagged pointer pointing to it. See Figure \ref{fig6}.

Clearly a necessary condition for such representation sharing is that
the two bodies have the same contents. Moreover, since a term body
always belongs to a single segment, the condition worked out for
absorption for two individual cells must hold for each pair of
corresponding body elements. We arrive at the following

\paragraph{Definition:} Term T1 can absorb term T2 if T1 is older than
T2, T1 == T2 and neither T1 nor T2 contain trailed cells.

Figure \ref{fig6} shows two bodies that fulfill the conditions.

Note that the example exhibits a situation we have not yet described:
a variable chain. Dereferencing must be stopped when a trailed cell is
found.

Note the similarity of the above analysis with the one for variable
shunting in \cite{VariableShunting}.

An algorithm that given two terms decides whether one can absorb the
other is now easily constructed. However, the naive use of this
algorithm would be very inefficient.

\subsection{ Properties of our notion of {\em can absorb}}
Before going to the implementation of representation sharing, it is
good to understand some properties of the {\em can absorb} relation:
the optimality (if any) of our algorithms depends crucially on those
properties.

It is clear that {\em can absorb} is not symmetric: a newer term
cannot absorb an older term in a different segment.
Neither is {\em can absorb} anti-symmetric: case {\bf a} in Section
\ref{individualcell} shows that.

We denote by {\em absorbed(x,y)} the result of letting term x absorb
term y, of course under the condition that x can absorb y.

An important part of our definition of {\em can absorb} is that the
terms do not contain trailed cells: it implies that a candidate term
for absorbing or being absorbed can be recognized without knowing the
other term, i.e., one checks whether it contains trailed cells or not
and by keeping information about visited terms, one can assure that
this information about the terms can be gathered in time proportional
to the heap.

From the definition, it also follows that {\em can absorb} is
transitive:
\begin{Verbatim}[fontsize=\small, frame=single,samepage=true]
       (x can absorb y) && (y can absorb z) ==> x can absorb z
\end{Verbatim}

Finally, the absorption process is also associative, i.e.,
\begin{Verbatim}[fontsize=\small, frame=single,samepage=true]
       absorbed(absorbed(x,y),z) == absorbed(x,absorbed(y,z))
\end{Verbatim}
(of course under the condition that x can absorb y and y can absorb z).
This means that the order in which absorption takes place is
immaterial: the end result is the same.

Together, these properties allow for a basically linear sharing
algorithm, on condition that term hashing is perfect. With a less than
perfect hash function, the algorithm might need to traverse some terms
more than once.

\section{Implementation of Representation Sharing}\label{implementation}

We have taken hProlog as the platform for an implementation of
representation sharing. hProlog is based on the WAM
\cite{wam:hassan,DHWa83} with a few differences:
\begin{itemize}
\item 
the choicepoint stack and environment stack are not interleaved as in
the WAM, but separate stacks as in SICStus Prolog
\item 
free variables only reside on the heap; i.e., there are no
self-references in the environment stack, just as in Aquarius Prolog
\cite{Aquarius}
\item 
hProlog supports some more native types like char, string and bigint;
it also has attributed variables
\end{itemize}

hProlog employs a mark-and-copy type of garbage collector, with its roots
in \cite{BevemyrLindgren@PLILP-94}, and it preserves segment order as
described in \cite{VandeginsteSagonasDemoenPADL2002}. Most other
systems use a sliding collector based on
\cite{SicstusGarbage@CACM-88}.  hProlog does not implement variable
shunting.

hProlog is a direct descendant of dProlog \cite{wamvariations}. Its 
purpose is to offer a platform for experiments in WAM-like Prolog
implementation. Its high performance gives the experiments an extra
dimension of credibility.

The implementation uses two data structures: they can be seen in Figure
\ref{fig:pic1}. We name them {\em cached\_hash} table and {\em
hashed\_terms} table. Together they form the sharer tables.
\begin{itemize}
\item 
cached\_hash: this is an array the size of the WAM heap (or global stack)
and can be though of as parallel to the heap; its entries contain
information about the corresponding heap cells; the information is one
of the following three:
\begin{itemize}
\item 
{\bf no-info}: the corresponding heap cell has not been {\em treated}
yet
\item 
{\bf impossible}: the corresponding heap cell cannot participate in
representation sharing; see Section \ref{commentscode} for more on
this
\item 
a pointer to the hashed\_terms table: the corresponding heap cell has
been treated, and its sharing information can be found by following
the pointer
\end{itemize}

\item 
hashed\_terms: this data structure contains records with two fields:
{\em hashvalue} and {\em term}; suppose a pointer in the cached\_hash
points to a record in the hashed\_terms, and the corresponding heap
cell A is the entry point of term TermA, then
\begin{itemize}
\item 
the {\em hashvalue} field in the record is the hash value of term TermA
\item 
the {\em term} field in the record is a pointer to a
heap cell B that is the entry point of a term TermB that can absorb
TermA (provided A and B are the not same cell); our implementation
makes sure that the heap cell B is as old as possible, i.e., B is equal
to A or older than A
\end{itemize}

\end{itemize}

Treating a heap cell consists in filling out the corresponding cell in
the cached\_hash table and possibly the hashed\_terms table.

The implementation of the hashed\_terms table is actually as a hash
table: the hashvalue of a term modulo the size of the hash table is
used for determining the place in the hashed\_terms, and a linked list
of buckets is used to resolve collisions. Many other implementations
of this hashed\_terms table would be fine as well.

Our first description of the algorithm only tries to introduce sharing
between structures (not lists). Therefore, for now, hashed\_terms pointers
can only appear in cached\_hash cells corresponding to a heap cell
containing a functor descriptor.

The main algorithm consists of two phases:
\begin{itemize}
\item 
{\bf build}: it builds the cached\_hash and hashed\_terms tables;
during this phase nothing is changed to the heap; this phase {\em
treats} all heap cells
\item 
{\bf absorb}: it performs all absorption possible by using the cached\_hash
and hashed\_terms tables
\end{itemize}

In the algorithms below, we use beginheap and endheap for the pointers
to the first (oldest) cell in the heap and the last (newest). We
assume no cell is trailed, and come back to this point later.

\subsection{Phase I: building the cached\_hash and hashed\_terms tables}\label{fase1}

\begin{sloppypar}
The build phase performs the action {\em compute\_hash} for each cell
in the heap: the corresponding cell in the cached\_hash is set to
either {\em impossible} or to a pointer to the hashed\_terms. The
function {\em compute\_hash} is always called with a tagged term as
argument. In the code below, we use {\em STRUCT} as the tag of a
pointer pointing to the functor cell of a compound term. In figures,
this tag shows simply as {\em S}. The function call {\em
tag(p,STRUCT)} returns such a STRUCT tagged pointer; the function {\em
untag} has the oposite effect. A function call like {\em tag(term)}
returns the tag of its argument.
\end{sloppypar}

Note that the following code ignores certain issues like checking
whether a cell is trailed, and LISTS. We deal with them in Section
\ref{commentscode}.

\begin{Verbatim}[fontsize=\small, frame=single,samepage=true]
foreach p in [beginheap, endheap] && is_functor(*p)
     compute_hash(tag(p,STRUCT));  // ignore return value

int compute_hash(p)
{
  deref(p);
  switch tag(p)
  {
    case FREE:
    case ATOMIC:
         return(p);

    case STRUCT:
         p = untag(p,STRUCT);
         if (already_computed(p)) return(already_computed_hash(p));
         hashvalue = *p;
         foreach argument of structure p do
                 hashvalue += compute_hash(argument);
         save_hash(hashvalue,p);
         return(hashvalue);
  }
}
\end{Verbatim}

The particular hash value computed above is not relevant for our
discussion: in practice, there are better (more complicated) ways to
compute hash values of terms.

The function call {\em already\_computed(p)} checks whether the
corresponding element in the cached\_hash table points to the
hashed\_terms table.  {\em already\_computed\_hash(p)} returns the
hash value previously computed (for the term starting at p) from the
hashed\_terms entry corresponding to p: in this way, re-computation
(and re-traversal of the same term) is avoided.

In the {\em save\_hash} function that follows, we have left out
collision handling: for the sake of the presentation, we assume
perfect hashing.

\begin{Verbatim}[fontsize=\small, frame=single,samepage=true]
save_hash(hashvalue,p)
{
  index = hashvalue % length(hashed_terms);

  cached_hash[p-beginheap] = hashed_terms + index;

  if (empty(hashed_terms[index]))
  { hashed_terms[index].term = p;
    hashed_terms[index].hashvalue = hashvalue;
    return;
  }

  // a non-empty entry might need to be adapted
  if newer(hashed_terms[index].term,p) hashed_terms[index].term = p;
}
\end{Verbatim}

The last line in save\_hash makes sure that the term pointed at in an
hashed\_terms entry is as old as possible. The reason is that it is
safe to let an older term absorb a younger one.

\begin{figure}[h]
\begin{centering}
\subfigure[After treating middle f(a,b)]{{\epsfig{file=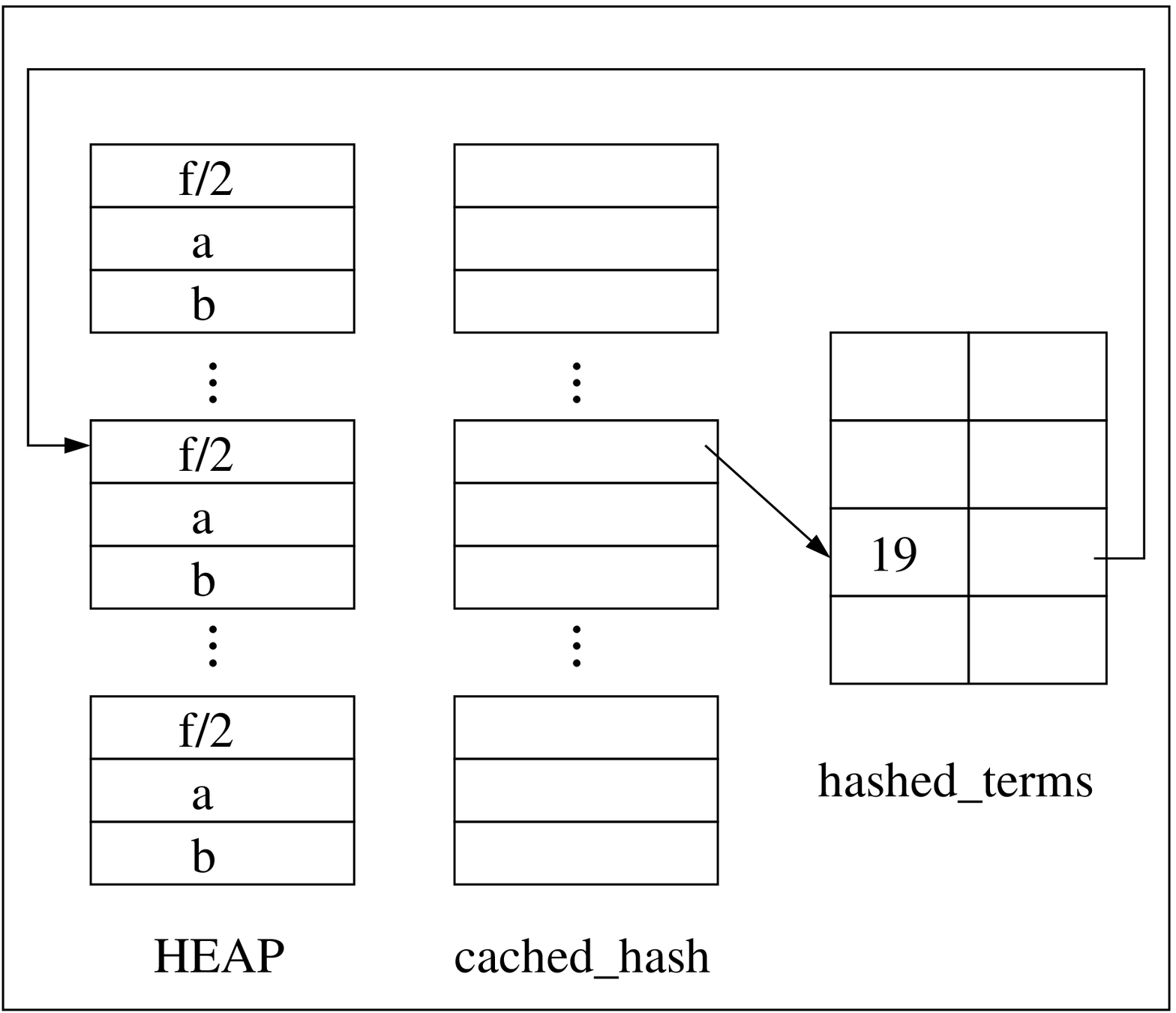,width=0.32\textwidth}} \label{implem2}}
\subfigure[After treating younger f(a,b)]{{\epsfig{file=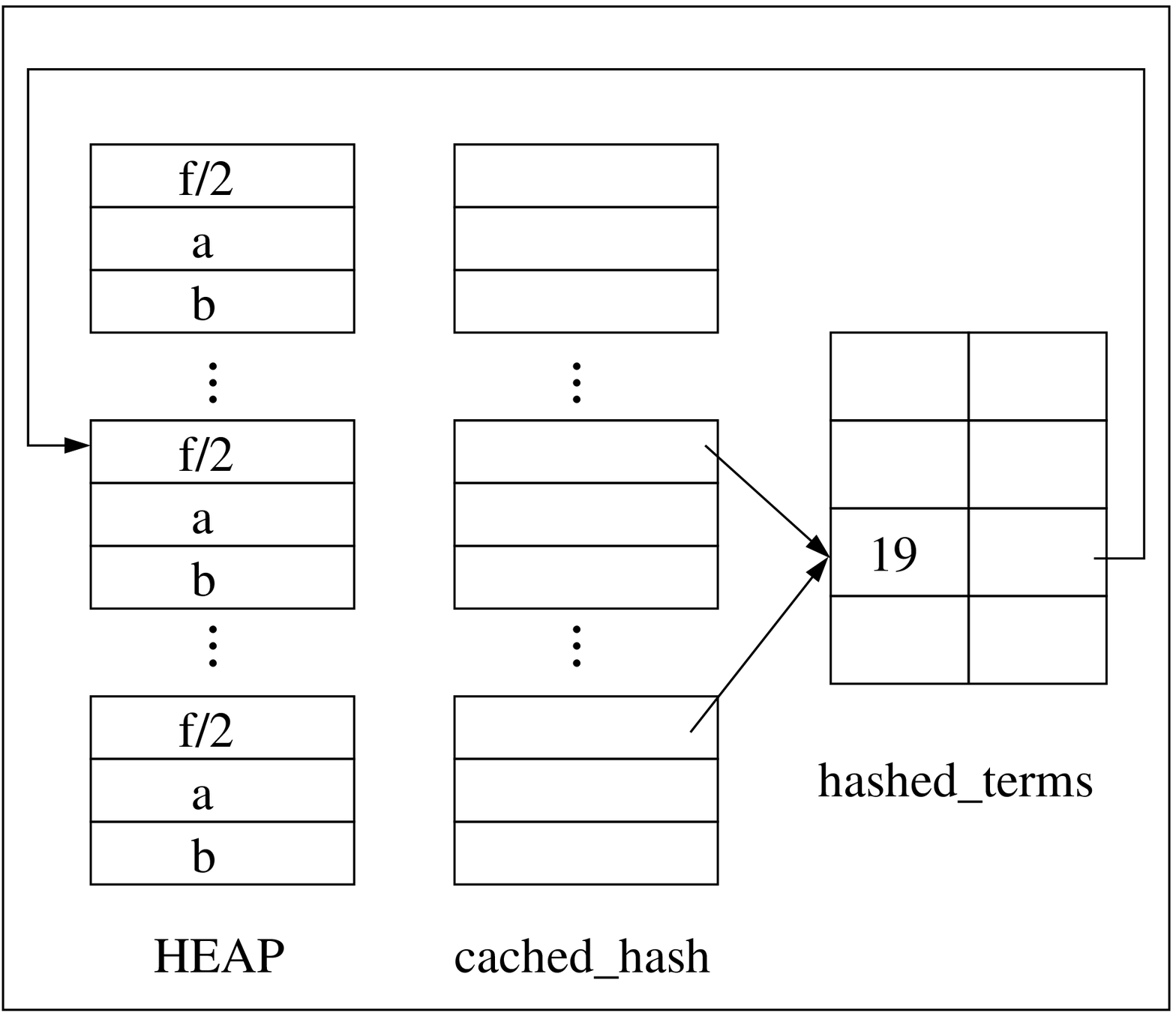,width=0.32\textwidth}} \label{implem3}}
\subfigure[After treating older f(a,b)]{{\epsfig{file=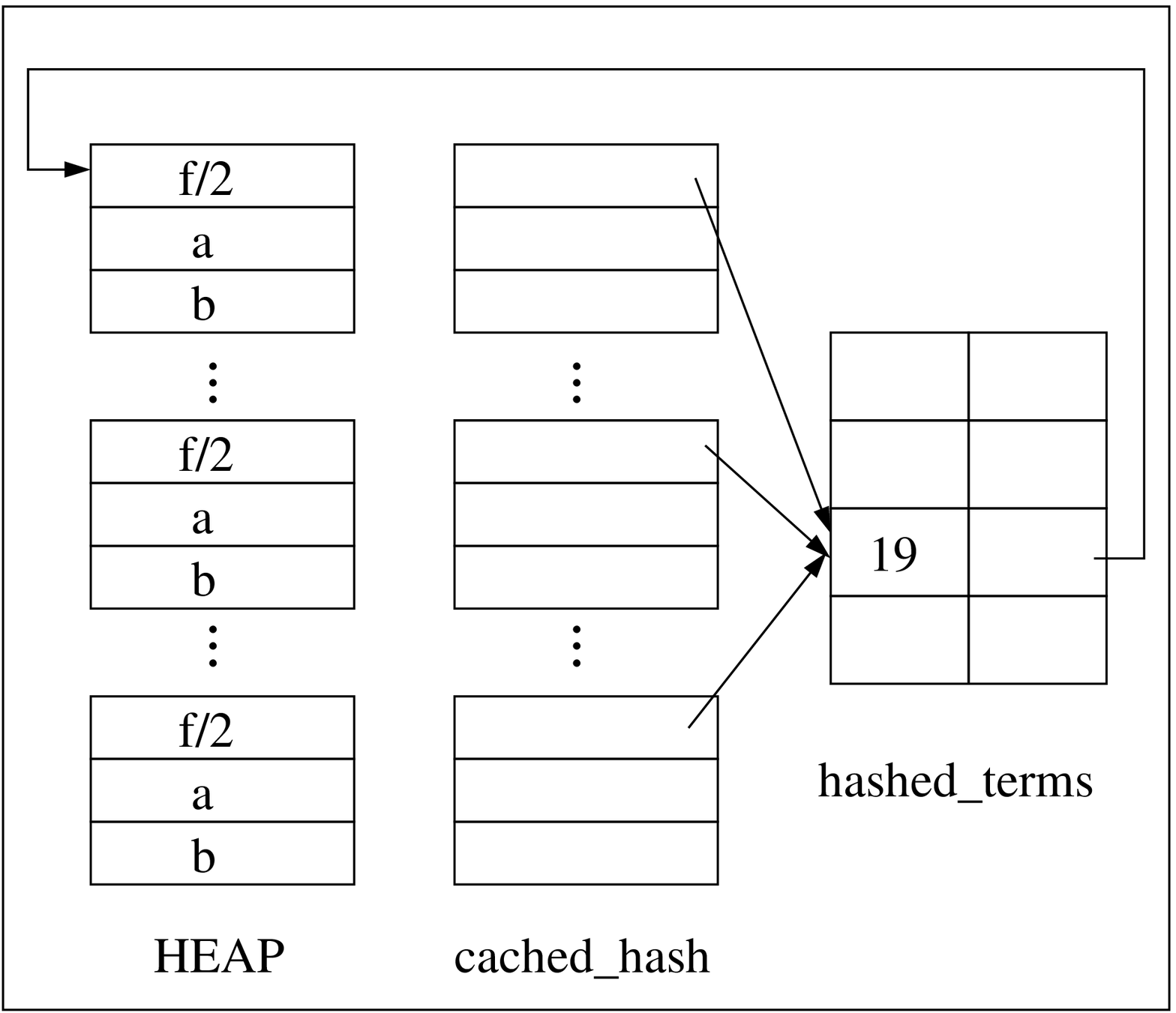,width=0.32\textwidth}} \label{implem4}}
\caption{Three identical terms are treated during the build phase}
\label{fig:pic1}
\end{centering}
\end{figure}

Figure \ref{fig:pic1} shows how three equal terms are treated by
compute\_hash and the effect thereof on the cached\_hash and hashed\_terms tables.

\subsection{Phase II: Absorbing}

The absorb phase performs the actual representation sharing: an
S-tagged pointer is redirected to the oldest term body that can absorb
it. The code is very simple:

\begin{Verbatim}[fontsize=\small, frame=single,samepage=true]
foreach cell c in the heap
               in the local stack
               in the choicepoint stack
               in the argument registers do
       let p be the contents of c;
       if (tag(p) == STRUCT)
          {
            q = untag(p,STRUCT);
            if (cached_hash[q-beginheap] points to hashed_terms)
                   replace c by tag(cached_hash[q-beginheap]->term,STRUCT); 
          }
\end{Verbatim}

\begin{figure}[h]
\begin{centering}
\subfigure[Just before absorbing]{{\epsfig{file=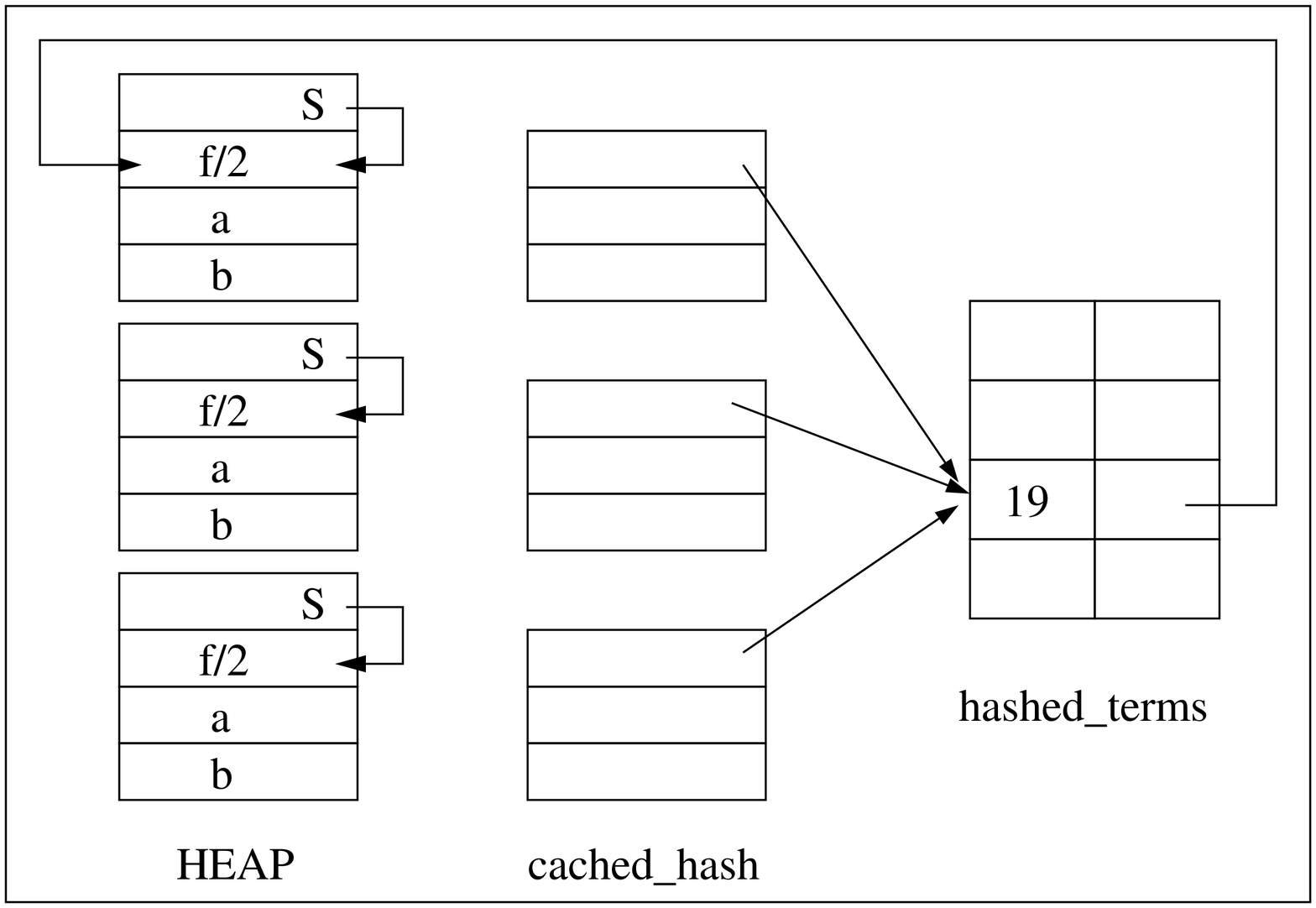,height=0.2\textheight}} \label{implem5}}
\subfigure[After absorbing]{{\epsfig{file=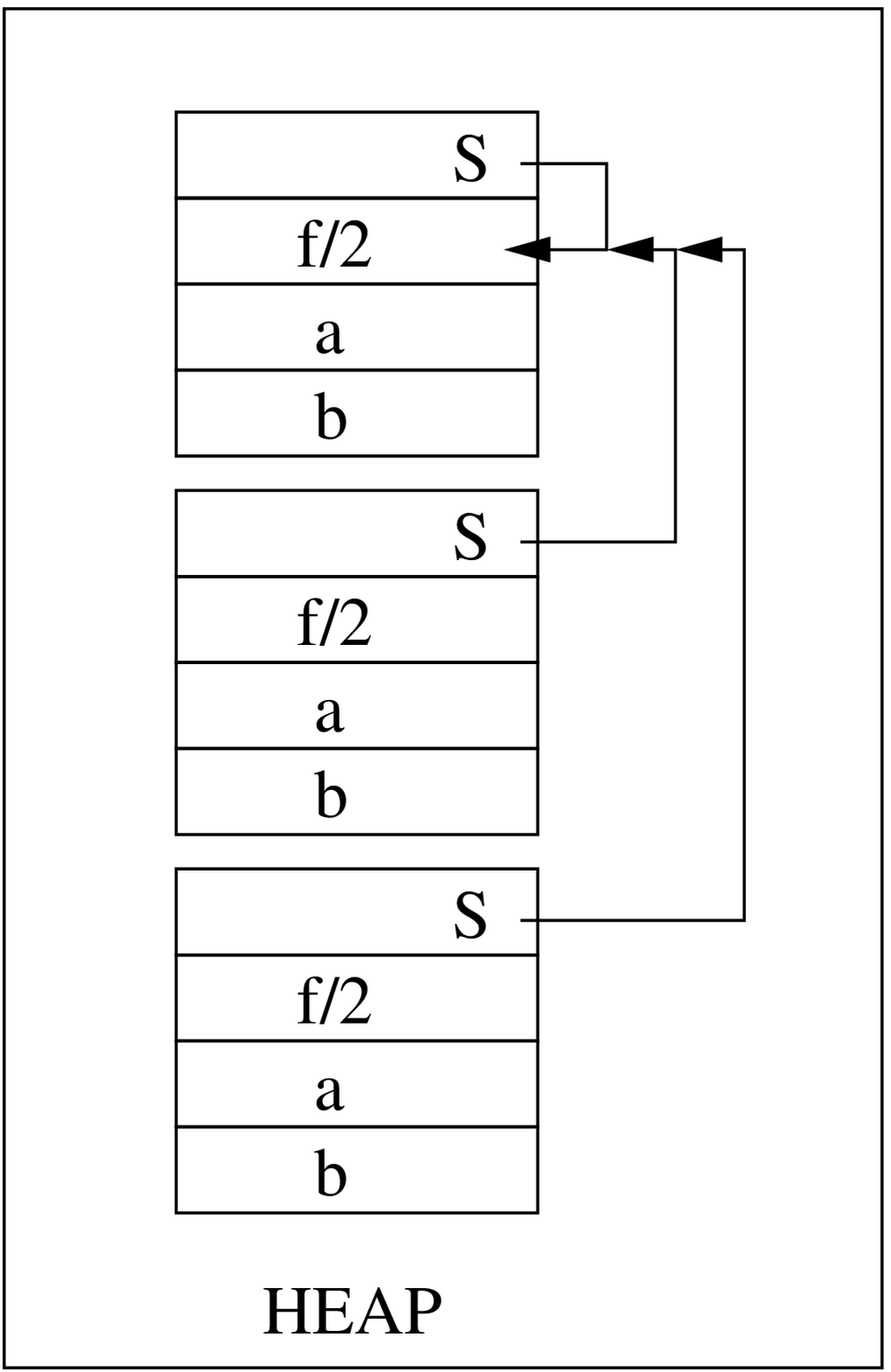,height=0.2\textheight}} \label{implem6}}
\subfigure[After one more GC]{{\epsfig{file=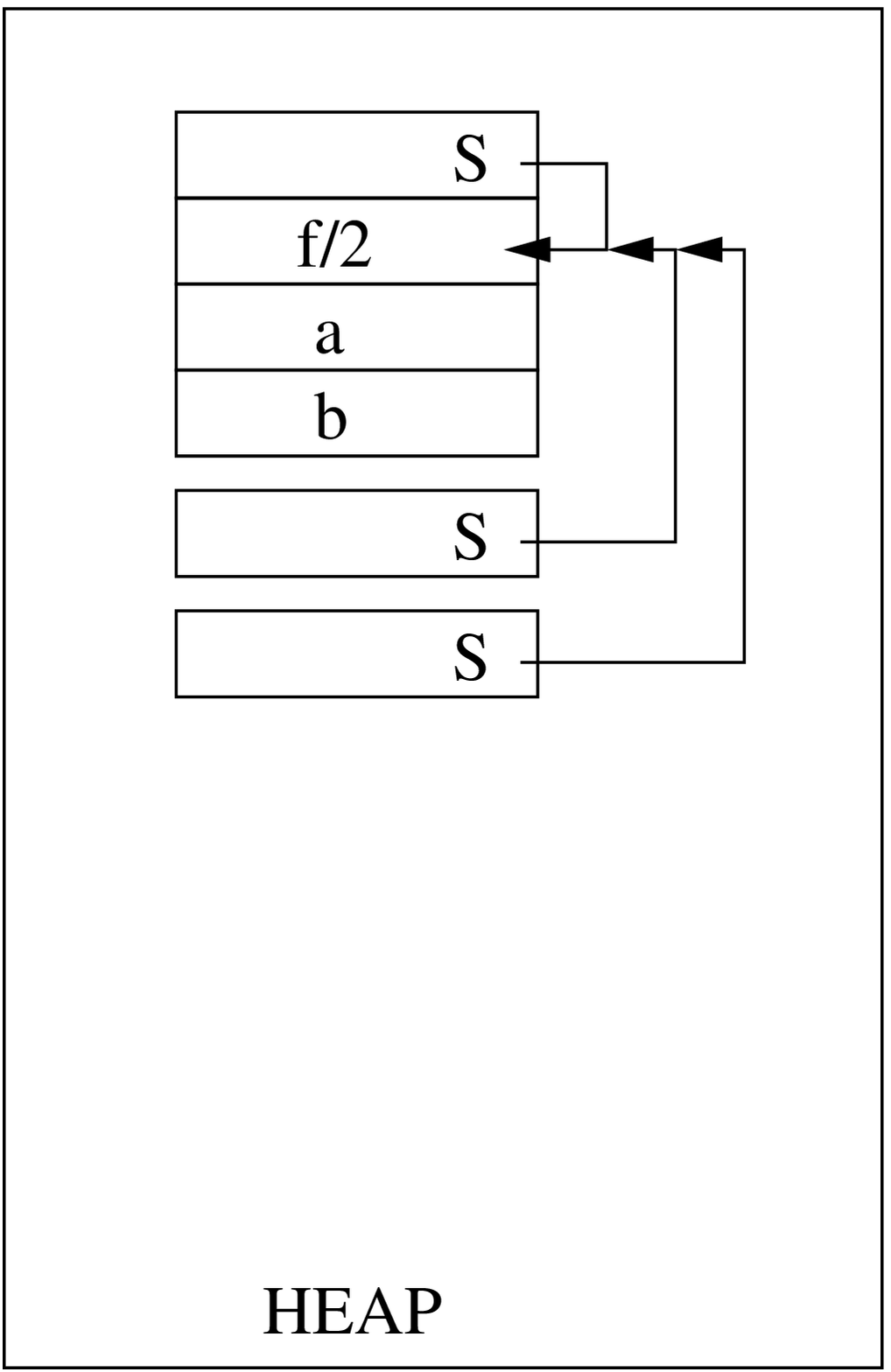,height=0.2\textheight}} \label{implem7}}
\caption{Absorption and GC in action}
\label{fig:pic2}
\end{centering}
\end{figure}

Figure \ref{fig:pic2} shows how the older term absorbs the two
identical younger terms.

\subsection{Comments on the Code}\label{commentscode}

The code in Section \ref{fase1} ignores certain issues:

\begin{itemize}
\item
{\em checking whether a heap cell is trailed}: during the
initialization of the build phase, the cached\_hash table entries
corresponding to trailed heap entries are initialized to {\em
  impossible}; this requires traversing the trail once and it makes
checking whether a cell is trailed constant time; the checks whether a
heap cell is trailed are required during the dereferencing loop; when
a trailed cell is encountered, the computation of the hash value is
stopped and the corresponding cached\_hash table entries of the term
containing the trailed cell are also set to {\em impossible}
\item 
{\em other datatypes}: the code takes into account only non-list
structured terms, atoms and variables; it is easy to extend it to
other types that occupy a single cell; for other {\em atomic} types
(real, string, bigint) we have followed the same principle as for
non-list structured terms: those types are implemented roughly like
such terms, i.e., with a tagged pointer to a header on the heap which
is followed by the actual value that can span several heap cells; for
lists, we have a different solution: see Section \ref{doinglists}
\item 
{\em foreach}: our implementation uses a linear scan for the {\em
  foreach} constructs: this is possible for all the stacks in hProlog;
if this is not the case, one can traverse the live data starting from
the root set as the garbage collector does (e.g., during its marking
phase)
\end{itemize}

The code implementing the above is less than 700 lines of plain C that
reuses very little previously existing code.

Note that in the context of our copying collector, the extra space
needed for representation sharing is just the hashed\_terms table: the
cached\_hash table has exactly the same size as the collector needs
for performing its collector duties.

\subsection{Representation Sharing of Lists}\label{doinglists}

\begin{figure}[h]
\begin{centering}
\subfigure[Just before absorbing]{{\epsfig{file=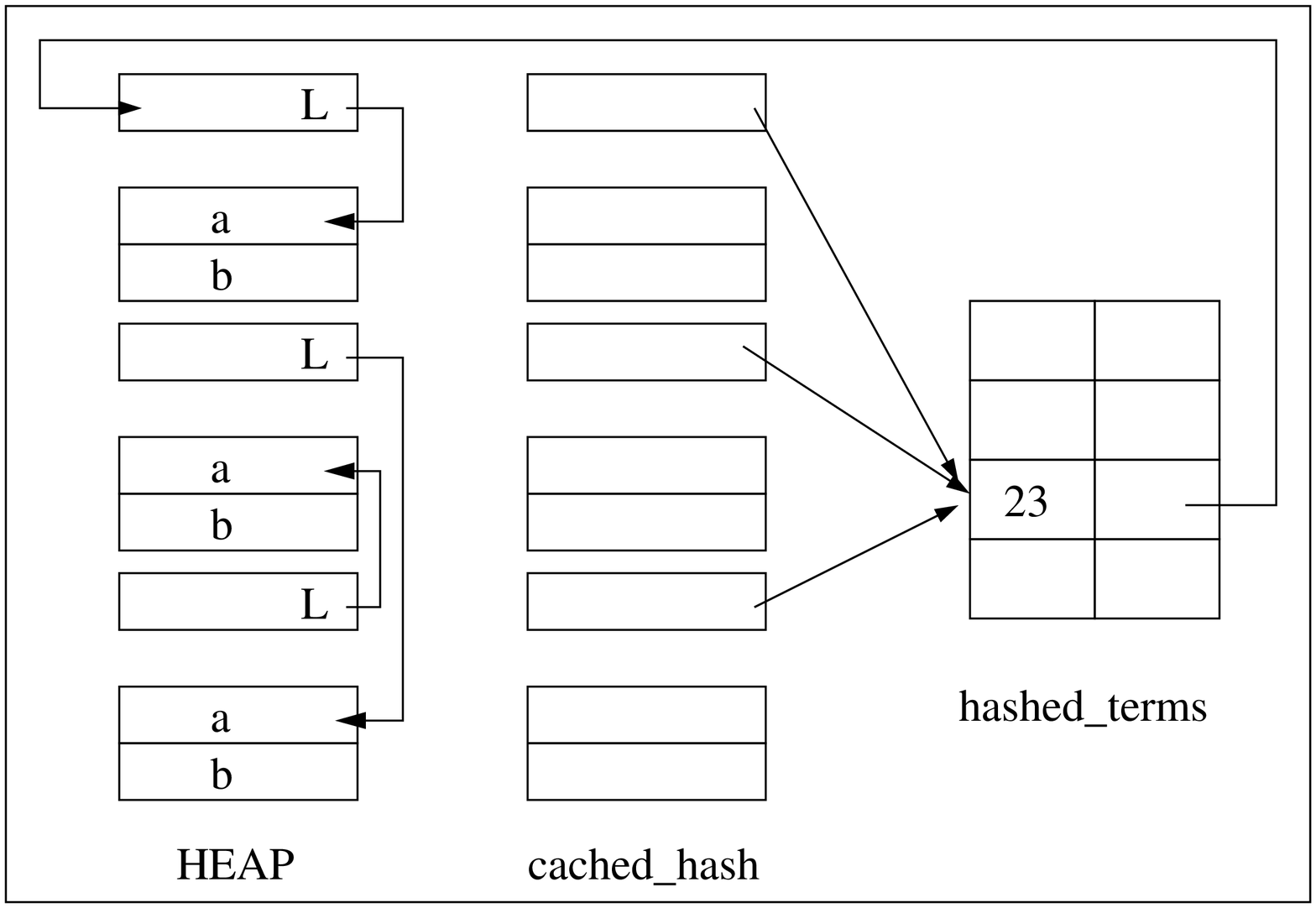,height=0.2\textheight}} \label{implem8}}
\subfigure[After absorbing]{{\epsfig{file=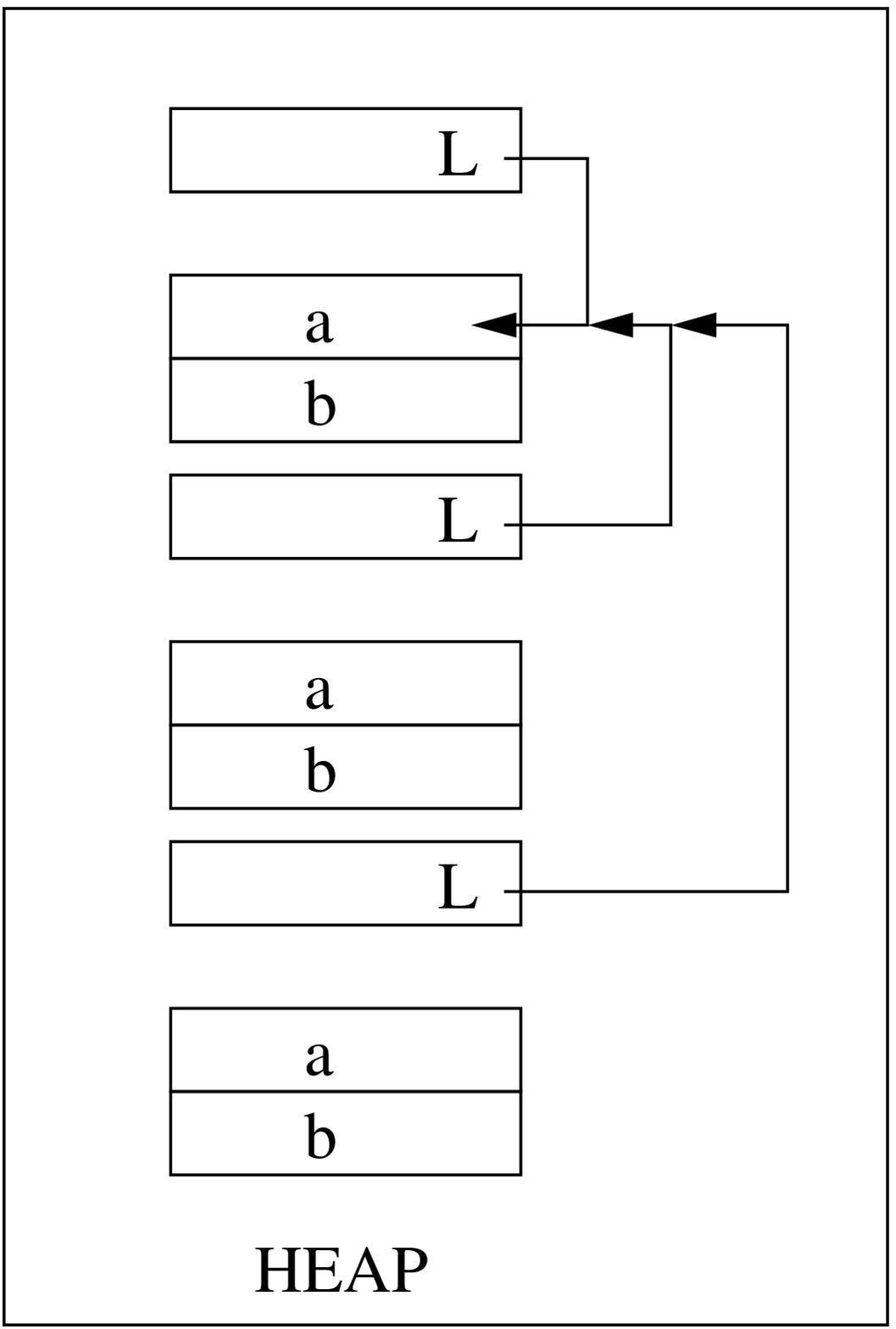,height=0.2\textheight}} \label{implem9}}
\subfigure[After one more GC]{{\epsfig{file=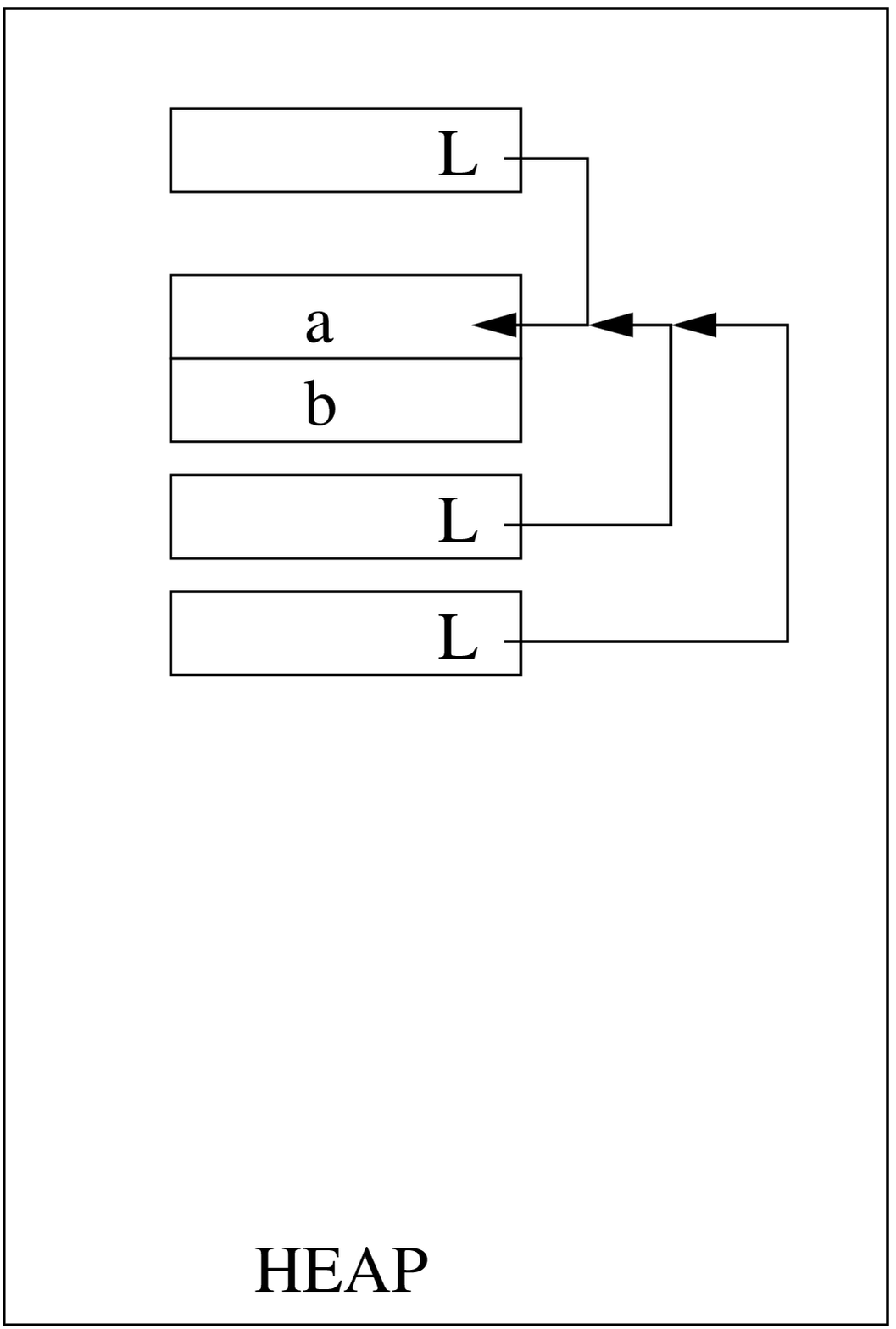,height=0.2\textheight}} \label{implem10}}
\caption{Absorption and GC in action}
\label{fig:pic3}
\end{centering}
\end{figure}

In the WAM, lists have no header like other compound terms. A list is
represented by an L-tagged pointer to two consecutive heap cells
containing the first element of the list and its tail
respectively. Clearly, we cannot deal with lists as in the previous
algorithm. The change is however small: we keep the hashed\_terms
pointer in the cell corresponding to the list-pointer.  Figure
\ref{fig:pic3} shows an example with just lists.

\begin{sloppypar}
Note that functor cells can only appear on the heap, while
list pointers can occur also in environments, choicepoints, and the
argument registers. As a result, with just a hashed\_terms pointer
array parallel to the heap, some representation sharing in the other
stacks can get lost for lists. A similar hashed\_terms pointer array
parallel to the other stacks can solve this problem: our
implementation does not do that. Another solution consists in using
the cell of the first element of a list for keeping the corresponding
hashed\_terms information. We have not explored that alternative.
\end{sloppypar}

\subsection{When to run the Sharer}

It seems obvious that the sharer must be run either during GC, or just
after GC. Our sharer can be adapted to run during GC most easily when
the GC starts with a marking phase: the build phase of the sharer can
indeed be integrated in the marking phase of the collector. The absorb
phase can be run before the next GC phase, or be integrated with
it. That would lead to a (mark+build)\&(copy+absorb) collector for
hProlog. In a sliding GC context, this would become
(mark+build)\&(compact+absorb).

Still, we choose from the beginning to run the sharer as an
independent module that could actually be run at any time. Just after
GC seems the best, because at that moment, the heap has minimal
size. We name that policy {\em after GC}.

There is one snag in this: the space freed by the sharer cannot be
used immediately, and the beneficial effect of the sharer can be seen
only after the next GC. Therefore, it feels like immediately after
the sharer, another GC should be done. We name that policy {\em
between GC}.

We have therefore added an option to hProlog:
\begin{itemize}
\item 
-r0: no sharing
\item 
-r1: sharer with policy {\em after GC}
\item 
-r2: sharer with policy {\em between GC}
\end{itemize}

Note that the absorb phase could estimate the amount of space it has
freed, and the decision to switch from one policy to the other could
be based on that.

\section{The Benchmarks and the Results}\label{sharerbenchmarks}

Since \cite{appelhashconsinggc} is closest to our representation
sharing, we are inclined to use the same benchmarks. However,
\cite{appelhashconsinggc} shows overall very little impact of {\em
hash-consing} and unfortunately, the benchmarks were not analyzed so
as to explain why hash-consing is not effective on them. On the other
hand, one cannot a priori assume that our sharer will show the same
behavior, because of the differences between our respective
implementations, and even the language:

\begin{itemize}
\item 
hProlog only performs major collections, while SML/NJ has a
generational collector (with two generations)
\item 
our sharer does not alter the representation of terms, while
\cite{appelhashconsinggc} performs hash-consing (which entails a
representation change) on the old generation only
\item 
SML/NJ is a deterministic language and a boolean SML/NJ function is like
a semi-det predicate in Prolog; however, in a typical Prolog
implementation, the data it creates is (on failure) backtracked over
in Prolog and the WAM recovers its space: this can have a huge impact
on some benchmarks (the mandelbrot benchmark is an example)
\item
in \cite{appelhashconsinggc} hash-consing was inseparably tied to the
(generational) collector; in contrast, we have explicitly aimed at
keeping the collector and the sharer separated (we argue why in
Section \ref{generalities}); this has an impact on the efficiency of
the sharing process
\end{itemize}

So it seems worthwhile to redo some of the benchmarks of
\cite{appelhashconsinggc}. The following section describes those
benchmarks as well as some others not appearing in
\cite{appelhashconsinggc}.

\subsection{The Benchmarks}\label{benchmarks}

\subsubsection{Boyer}\label{boyer}

Boyer is a famous benchmark initially conceived by R. Gabriel for
Lisp, and later used in other functional and logic
contexts. Essentially, it rewrites a term to a canonical form. Boyer
has been the subject of many studies, and in particular for proving
that it is not a good benchmark: see for instance
\cite{bakerwarpspeed}. Anyway, in \cite{appelhashconsinggc}, this
benchmark shows the best results for hash-consing. The inherent
reason is that terms are rewritten to a canonical form and thus
many initially different terms end up the same. We measured that the
final result of the rewriting process needs 39\,834 heap cells without
representation sharing, and only about 200 with representation
sharing.

This makes boyer close to an optimal benchmark for showing the
effectiveness of representation sharing.

Note that the boyer benchmark also benefits a lot from tabling
\cite{ChWa96}. This means that repeated computations are going on,
which explains also the high amount of representation
sharing. However, while tabling does avoid the repetition of duplicate
computations, as usually implemented, it does not avoid the creation
of duplicate terms on the heap. It is possible to add to the tries
enough info so that ground terms need be copied only once to the heap
as long as this copy is not backtracked over.

\subsubsection{Life}\label{life}

\cite{appelhashconsinggc} also uses the well known {\em Game of Life}
as a benchmark. We have written a version in Prolog following the
ideas of Chris Reade \cite{ChrisReade}, just as
\cite{appelhashconsinggc} did. A (live) cell is represented as a tuple
in coordinate form (X,Y). A generation is a list of live cells. The
program keeps a list of the first 1000 generations, starting from 
the {\em The Weekender}\footnote{See
http://fano.ics.uci.edu/ca/rules/b3s23/g10.html} which is a glider,
\begin{wrapfigure}{r}{.35\textwidth}
\begin{center}\includegraphics[%
  width=0.35\textwidth,
  keepaspectratio]{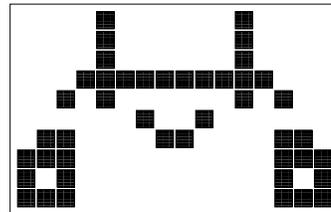}\end{center}
\caption{The Weekender}
\end{wrapfigure}
i.e., a pattern that repeats itself after a few generations (7 in this
case) translated a few cells (2 in this case). If just the most recent
generation is kept alive, one expects little from running the sharer
immediately after a major collection, as the just rewritten generation
is garbage. Our benchmark still shows some 50\% memory improvement,
because it keeps all computed generations in a list, so that the
existing overlap between generations is shared.

\cite{appelhashconsinggc} shows little gain from hash consing for this
benchmark, but we could not retrieve the initial generation(s) on
which the benchmark was run.

\subsubsection{Mandelbrot}

This benchmark was also used in \cite{appelhashconsinggc}: it computes
(actually outputs) a bitmap of a Mandelbrot set of a given
dimension. Since the output does not play a role in the heap usage, we
have removed the code for the output. We took the version from the
{\em Computer Language Benchmarks Game}
(http://shootout.alioth.debian.org/) written for Mercury and based on
a version by Glendon Holst. Mandelbrot uses quite a bit of heap and as
such appears a good memory benchmark. However, one can see quickly
that literally {\bf all} memory used by mandelbrot is by floating
point numbers: computed floating point numbers have the tendency to be
different and therefore representation sharing might not have much
effect. We have indeed checked that half of the generated floating
point numbers are unique during the benchmark.

Almost all the floating point numbers are generated during a ground
call to {\em mandel/5}, a semidet predicate called as the condition in an
if-then-else as follows:
\begin{Verbatim}[fontsize=\small, frame=single,samepage=true]
            (mandel(Height, Width, Y, X, 50) ->
                   ByteOut1 is (ByteOut0 << 1)
            ;
                   ByteOut1 is (ByteOut0 << 1) \/ 0x1
            )
\end{Verbatim}

In the setting of \cite{appelhashconsinggc} (generational collection +
hash-consing) the Mandelbrot benchmark has the following
characteristic: if the garbage collector runs during the test
({\em mandel/5}) then a few floats are copied to the older generation,
otherwise, no float from the new generation survives the collection.
So, not even all computed floats end up in the zone subject to
hash-consing.

In our setting (only major collections + representation sharing),
at each collection, only some floats in the test are alive. Exactly at
that moment, the chance for duplicates is very small.

The effect of hash-consing or representation sharing is expected to be
very small for the mandelbrot benchmark.

Our test runs of the mandelbrot benchmark indeed show zero gain from
representation sharing.

\subsubsection{One more classical Prolog benchmark: {\em tsp}} 

We were unable to retrieve more benchmarks from
\cite{appelhashconsinggc}, so we tried different benchmarks from the
established general Prolog benchmark suite. None showed any benefit
from representation sharing. We report only on {\em tsp}: just like
{\em mandelbrot} and the other benchmarks showing no benefit, it is
mainly good for showing the overhead of the useless sharer.

\subsubsection{{\em blid/1}}\label{blid}

\begin{sloppypar}
The next program was altered slightly from what Ulrich Neumerkel
posted in comp.lang.prolog; it appears also in his Diplomarbeit
\cite{neumerkeldiplomarbeit}.
\end{sloppypar}
\begin{Verbatim}[fontsize=\small, frame=single,samepage=true]
blid(N) :-                           blam([]).
       length(L, N),                 blam([L|L]) :-
       blam(L),                             blam(L).
       id(L,K),                     
       use(K).                       id([], []).
                                     id([L1|R1], [L2|R2]) :-
                                            id(L1,L2), % L1 = L2
use(_).                                     id(R1,R2). % R1 = R2
\end{Verbatim}
His question was {\em Are there systems, that execute a goal blid(N)
in space proportional to N?  Say blid(24)}. At first we expected that
with our representation sharing, space would be indeed linear in
N. However, the expansion policy and order in which events (garbage
collection and representation sharing) take place is also crucial. 
\begin{itemize}
\item 
with the {\em after GC} policy, the following happens:
\begin{itemize}
\item[a1:] 
the first GC finds that 99\% (or more) of the data is live, and
decides to expand the heap
\item[a2:] 
the sharer shares most data
\item[a3:] 
the next triggered GC finds that about half of the heap is live, so
does not expand
\item[a4:]
the following sharer shares most of the data
\item[a5:] 
points a3 and a4 are repeated
\end{itemize}

\item 
with the {\em between GC} policy, the following happens:
\begin{itemize}
\item[b1:] 
the first GC finds that 99\% (or more) of the data is live, and
decides to expand the heap
\item[b2:] 
the sharer shares most data
\item[b3:] 
the second GC collects almost all data
\item[b4:]
points b1, b2 and b3 are repeated
\end{itemize}

\end{itemize}

The first GC in a1 and b1 is triggered by lack of space, the second GC
(in b3) is there by policy. A GC can decide to expand the heap (in
hProlog when the occupancy is more than 75\%: this is known after
marking). So one sees that in the case of the {\em between GC} policy,
the heap is repeatedly expanded, even though the program could run in
constant space (with the aid of the sharer). With the {\em after GC},
we do not get into this repeated expansion.

If hProlog also had a heap shrinking policy, the {\em between
GC} policy would after its second collection shrink the heap, and this
would amount to almost the same effect as the {\em after GC} policy.

This shows that the combination of a reasonable heap expansion policy
and a reasonable sharer policy can result in an overall bad policy.
More work could be done on this.

Note that in its original form, {\em id/2} also contains the two commented
out unifications, and that with unification factoring these would also
introduce the sharing needed to run in O(N) heap (always with the aid
of GC of course).

\subsubsection{Four Applications}

The next four benchmarks provide some insight in what to expect from
the sharer in a some typical applications of Prolog: there is little
impact on memory and performance.

\paragraph{Tree Learner.}

This realistic benchmark consists of a {\em best-first relational
regression tree learner} written by Bernd Gutmann \cite{Gutmann}. The
program is about 900 LOC. It works on a data set of 350K facts.

\paragraph{Emul.}

\begin{sloppypar}
Emul is a BAM emulator \cite{Aquarius} written by Peter Van Roy in
Prolog. The benchmark consists in executing the BAM code for the
famous {\em SEND+MORE=MONEY} problem. It is about 1K LOC. 
\end{sloppypar}

\paragraph{An XSB compiler.}

xsbcomp is an old version of the XSB compiler \cite{SaSW94} and a run
of the benchmark consists in compiling itself. The XSB compiler is
about 5K LOC and also uses the hProlog or SICStus Prolog reader which
are also in Prolog.

\paragraph{The hProlog compiler.}

In this benchmark, the hProlog compiler compiles itself. It uses {\em
setarg/3} heavily, so we need a {\em mutable/1} (see Section
\ref{mutable}) declaration for 7 functors. This benchmark cannot be
run by SICStus. The hProlog compiler (which is a version of the hipP
compiler \cite{querypacks} written by Henk Vandecasteele) plus all
other code it needs (reader, optimizer ...) totals more than 10K LOC.

\subsubsection{Worst and best Case}

It is not clear what the best and worst case for our sharer is: if the
heap were just one huge flat term (say of the form f(1,2,3,...)) then
only one hash value would have to be saved in the hashed\_terms table,
and in some sense that is both best and worst, because the least time
is lost in collisions etc, but also no sharing can be performed. So we
choose the following as best-versus-worst case: a large complete
binary tree in which every node is of the form {\em
node(tree,tree,number)}. In what we consider the best case, the number
is always the same (and thus resembles a bit the blid data structure
in Section \ref{blid}). This leads to a very sparse hashed\_terms
table, and a large amount of sharing. In the worst case, the number is
different in all nodes: as a result, the hashed\_terms becomes quite
full, and no sharing is possible at all. The main reason for this
benchmark is to find out how the build and the absorb phase contribute
to the total time of the sharer.

\subsection{The Benchmark Results for Representation Sharing}

The results are shown in Table \ref{sharertimespace}. Time is in
milliseconds. Space is in Mib or Kib as indicated in the table.

The first two columns denote the benchmark and system used (with the
sharing option for hProlog). Then follow the total time taken by heap
garbage collection (including the stack shifter), the total time taken
by the sharing module, the total execution time and the number of
garbage collections. Then follow four columns related to space: the
initial heap size, the final heap size and the amount of space
collected by the garbage collections are given in megabytes. Finally,
there is the heap high water mark at the end of the benchmark given in
KiB instead of Mib because the figures vary widely. It measures the
size of the result computed by the benchmark and it includes a small
system specific overhead from the toplevel: for {\em mandelbrot}, the
figure is just that overhead.

\begin{table}[!]
\begin{center}
\footnotesize
\begin{tabular}{|r|r||r|r|r||r|r|r|r|r|} \hline
bench      & system        & gc     & share  &  total  &\#gc   & initial  & final     &collected&at end \\
           &               & time   & time   & runtime &       & heap     & heap      &         &         \\
           &               & msecs  & msecs  &  msecs  &       & Mib       & Mib        &  Mib     &  Kib  \\
\hline
\hline
boyer & hProlog -r0 & 920 & 0 & 3130 & 19 & 6.10 & 24.42 & 96.75 & 6953.96 \\ 
boyer & hProlog -r1 & 280 & 380 & 2820 & 24 & 6.10 & 6.10 & 111.60 & 287.76 \\ 
boyer & hProlog -r2 & 260 & 380 & 2750 & 36 & 6.10 & 6.10 & 109.74 & 0.91 \\ 
boyer & SICStus  & 1440 & - & 6170 & 83 & 6.10 & 26.01 & 240.76 & 6954.23 \\ 
\hline 

life & hProlog -r0 & 2570 & 0 & 28960 & 100 & 3.76 & 15.04 & 17.70 & 8896.39 \\ 
life & hProlog -r1 & 1400 & 3060 & 30860 & 100 & 3.76 & 7.52 & 22.23 & 4211.53 \\ 
life & hProlog -r2 & 2820 & 2910 & 32200 & 200 & 3.76 & 7.52 & 22.27 & 4164.63 \\ 
life & SICStus  & 4730 & - & 69250 & 100 & 3.76 & 9.91 & 22.20 & 8896.60 \\ 
\hline 

mandelbrot & hProlog -r0 & 10 & 0 & 29370 & 360 & 16.04 & 16.04 & 5774.14 & 0.09 \\ 
mandelbrot & hProlog -r1 & 50 & 0 & 29460 & 360 & 16.04 & 16.04 & 5774.14 & 0.09 \\ 
mandelbrot & hProlog -r2 & 20 & 10 & 29390 & 720 & 16.04 & 16.04 & 5774.14 & 0.09 \\ 
mandelbrot & SICStus  & 2090 & - & 219070 & 171 & 16.04 & 16.02 & 2656.25 & 0.36 \\ 
\hline 

tspgc & hProlog -r0 & 1190 & 0 & 47960 & 1625 & 3.76 & 3.76 & 5684.49 & 258.23 \\ 
tspgc & hProlog -r1 & 1070 & 1610 & 52030 & 1625 & 3.76 & 3.76 & 5684.49 & 258.23 \\ 
tspgc & hProlog -r2 & 2730 & 1770 & 55540 & 3250 & 3.76 & 3.76 & 5684.49 & 258.23 \\ 
tspgc & SICStus  & 3760 & - & 96430 & 1304 & 3.76 & 3.77 & 2867.55 & 258.43 \\ 
\hline 
\hline 

blid & hProlog -r0 & 1860 & 0 & 2480 & 6 & 3.76 & 240.64 & 0.01 & 131072.08 \\ 
blid & hProlog -r1 & 970 & 1910 & 3520 & 34 & 3.76 & 7.52 & 123.94 & 301.76 \\ 
blid & hProlog -r2 & 860 & 1750 & 3250 & 10 & 3.76 & 120.32 & 116.52 & 0.26 \\ 
blid & SICStus  & 5660 & - & 7050 & 15 & 3.76 & 179.22 & 0.00 & 131072.33 \\ 
\hline 

worst & hProlog -r0 & 40 & 0 & 80 & 1 & 16.04 & 16.04 & 0.01 & 8192.06 \\ 
worst & hProlog -r1 & 70 & 110 & 240 & 1 & 16.04 & 16.04 & 0.01 & 8192.06 \\ 
worst & hProlog -r2 & 120 & 100 & 280 & 2 & 16.04 & 16.04 & 0.01 & 8192.06 \\ 
worst & SICStus  & 90 & - & 240 & 1 & 16.04 & 16.05 & 0.00 & 8192.33 \\ 
\hline 

best & hProlog -r0 & 40 & 0 & 80 & 1 & 16.04 & 16.04 & 0.07 & 8192.06 \\ 
best & hProlog -r1 & 70 & 100 & 220 & 1 & 16.04 & 16.04 & 0.01 & 0.37 \\ 
best & hProlog -r2 & 70 & 110 & 240 & 2 & 16.04 & 16.04 & 8.01 & 0.37 \\ 
best & SICStus  & 90 & - & 230 & 1 & 16.04 & 16.05 & 0.00 & 8192.33 \\ 
\hline 
\hline

treelearner & hProlog -r0 & 990 & 0 & 41380 & 88 & 35.19 & 35.19 & 2936.31 & 0.08 \\ 
treelearner & hProlog -r1 & 920 & 760 & 39900 & 88 & 35.19 & 35.19 & 2936.50 & 0.08 \\ 
treelearner & hProlog -r2 & 1980 & 750 & 40570 & 176 & 35.19 & 35.19 & 2936.75 & 0.08 \\ 
treelearner & SICStus  & 5340 & - & 95860 & 75 & 42.13 & 42.08 & 2947.34 & 0.35 \\ 
\hline 

emul & hProlog -r0 & 30 & 0 & 4610 & 70 & 3.76 & 3.76 & 258.28 & 0.08 \\ 
emul & hProlog -r1 & 60 & 20 & 4500 & 70 & 3.76 & 3.76 & 260.04 & 0.08 \\ 
emul & hProlog -r2 & 80 & 20 & 4310 & 140 & 3.76 & 3.76 & 260.08 & 0.08 \\ 
emul & SICStus  & 740 & - & 8330 & 90 & 3.76 & 3.72 & 318.77 & 0.36 \\ 
\hline 

xsbcomp & hProlog -r0 & 20 & 0 & 390 & 3 & 3.76 & 3.76 & 10.02 & 0.07 \\ 
xsbcomp & hProlog -r1 & 20 & 0 & 360 & 3 & 3.76 & 3.76 & 10.48 & 0.07 \\ 
xsbcomp & hProlog -r2 & 20 & 0 & 410 & 6 & 3.76 & 3.76 & 10.72 & 0.07 \\ 
xsbcomp & SICStus  & 30 & - & 850 & 4 & 3.76 & 3.77 & 12.96 & 0.35 \\ 
\hline 

dpcomp & hProlog -r0 & 190 & 0 & 850 & 13 & 3.81 & 3.81 & 29.76 & 0.07 \\ 
dpcomp & hProlog -r1 & 130 & 110 & 900 & 11 & 3.81 & 3.81 & 29.62 & 0.07 \\ 
dpcomp & hProlog -r2 & 240 & 120 & 1010 & 22 & 3.81 & 3.81 & 30.30 & 0.07 \\ 
\hline 

\end{tabular}

\end{center}
\caption{The Sharer and the Collector}\label{sharertimespace}
\end{table}
\begin{sloppypar}
Table \ref{sharertimespace} shows sometimes a large difference between
the memory consumption of SICStus Prolog and hProlog. Also the time
spent in garbage collection, and the number of collections can be very
different. The reason is that although both systems are based on the
WAM, they differ in a number of other design decision. In particular,
their heap expansion policy differs, their garbage collectors differ
(the SICStus Prolog one is generational and compacting, while the
hProlog one is non-generational and copying), they have a different
approach to floating point arithmetic, and hProlog does not allocate
free variables in the local stack.
\end{sloppypar}

In addition to the results in Table \ref{sharertimespace}, we can also
mention that the build phase takes between 8.3 (for blid) and 2.6 (for
worst) times as long as the absorb phase: the absorb phase is indeed
much simpler.

\subsection{Conclusions from the Benchmarks}

By and large our results confirm the findings of
\cite{appelhashconsinggc}: most benchmarks hardly benefit from
representation sharing, and sometimes the space and time performance
becomes worse. Apart from the artificial benchmark {\em blid/1}, only for boyer
do we find a much larger ---huge in fact--- benefit from representation
sharing than in \cite{appelhashconsinggc}. We have not been able to
pinpoint why: the benchmarks used in \cite{appelhashconsinggc} are not
even available anymore, let alone the {\em queries}. The fact that
generational collection retains terms longer than an
only-major-collections strategy might play a role. Still, our result
is in line with the (confluent) rewriting character of boyer.

The time taken by our implementation of representation sharing is
reasonable: the algorithm is linear in the size of the heap,
localstack, choicepointstack and trail, so complexity wise not worse
than an actual garbage collection. The traversal of the stacks is
however less complicated, since one does not need to take into account
the liveness of the locations anymore and less copying is going on.
In our application benchmarks, the sharer always takes less time than
the garbage collection. It is clear that a better policy, and
improvements to our implementation code, can make the sharer even more
efficient. Our sharer does not depend on the efficiency of the
underlying Prolog system, neither its garbage collector, so we feel it
is safe to say that our sharer can be implemented with the same (or
better) performance in other WAM-like systems.

\section{Variations, Extensions and related Issues}\label{variants}

\paragraph{\bf Unusual Sharing.}

\begin{figure}[h]
\begin{centering}
{\epsfig{file=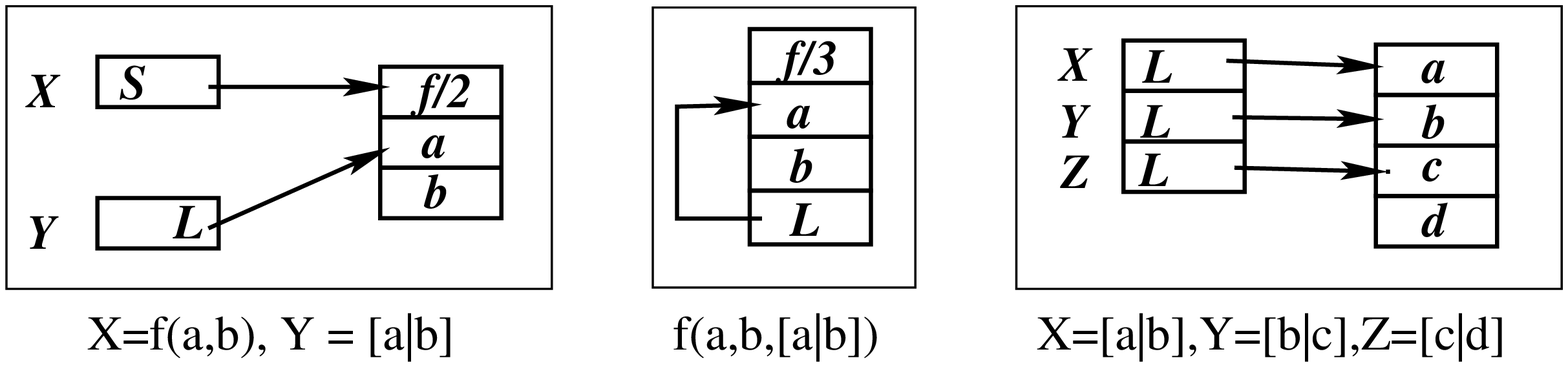,width=1.0\textwidth}}
\caption{Unusual representation sharing}
\label{fig:layout2}
\end{centering}
\end{figure}

In \cite{DemoenICLP2002fresh}, the rather unusual representation
sharings depicted in Figure \ref{fig:layout2} are described.

Our current representation sharing implementation does not achieve the
above sharings. Still, all ingredients are present and while the
expected gains are small, it is nice that the above unusual sharing
can be achieved in time linear in the size of the heap (assuming perfect
hashing).

\paragraph{\bf Cyclic Terms.}

\cite{appelhashconsinggc} deals with cyclic terms by excluding them
from hash-consing. It is easy to do the same in our implementation as
follows:
\begin{enumerate}
\item 
besides the special values {\bf no-info} and {\bf impossible}, 
cached\_hash entry can also have the value {\bf busy}
\item 
when a functor cell is visited for the first time, that corresponding
cached\_hash entry is set to {\bf busy}
\item 
when a functor cell is visited recursively, a check on the
corresponding cached\_hash entry detects that there is a cycle: the
field is set to {\bf impossible}
\item
as usual, when a term is visited completely, its corresponding field
is set to an appropriate value, i.e., {\bf impossible} or a pointer to
the hashed\_terms table
\end{enumerate}

However, one can do better: a variation of point 3 above yields a
procedure that can perform representation sharing also for cyclic
terms.
\begin{enumerate}
\item[3'.]
when a functor cell is visited recursively, a check on the
corresponding cached\_hash entry detects that there is a cycle: a
fixed value (say 17) is returned as the hash value of this term; the
corresponding cached\_hash entry is not updated at this time: this
happens when the visit has returned to the point where the entry was
set to {\bf busy}
\end{enumerate}

The procedure for testing equality of terms must also be adapted to
deal correctly with cycles: this is common practice now in most Prolog
systems.

Note that it does not matter which value is chosen in (3') above.
What matters is only that the hash value of terms that can share their
representation is the same. Still, our procedure can attach a
different hash value to cyclic terms that are equal (in the sense of
{\em ==/2}) and could share their representation. This results in no
representation sharing for those cyclic terms. As an example:
\begin{Verbatim}[fontsize=\small, frame=single,samepage=true]
        test :- X = f(1,f(1,X)), share, use(X).
\end{Verbatim}
does not result in the same heap representation as
\begin{Verbatim}[fontsize=\small, frame=single,samepage=true]
        test :- X = f(1,X), use(X).
\end{Verbatim}
The procedure based on minimization of finite automata described in
\cite{neumerkeldiplomarbeit} does.

\paragraph{\bf Mutable Terms.}\label{mutable}

Prolog systems supporting destructive update ---through {\em setarg/3},
mutable terms or for attributed variables--- often do this using a
trail in which each entry keeps the old value: clearly, these old
values can point to sharable terms and they can be updated accordingly
in the final absorb phase.

However, just as a ground mutable term must be copied by {\em copy\_term/2},
a mutable term itself is not allowed to absorb or be absorbed. This
means that mutable terms should be recognizable during the build
phase. In SICStus Prolog this is the case ({\em \$mutable/2} is reserved
for this), but not so in other systems (e.g., SWI Prolog, Yap, hProlog
...). In hProlog we have resolved that problem by introducing a
declaration: {\em :- mutable foo/3.} declares that the arguments of
any {\em foo/3} term can be destructively updated, and effectively prevents
sharing of {\em foo/3} terms. We use one bit in the functor table and the
overhead during the build phase is unnoticeable.
Note that the {\em :- mutable} declaration does not readily work
across modules.

\paragraph{\bf Cooperation between Collector and Sharing.}

We have implemented the representation sharing module independent of
the garbage collector module. The advantage is less dependency and
a higher potential that the sharer can be integrated in other
systems. The disadvantage is that some information that the garbage
collector has computed, needs to be recomputed by the sharing
module. For instance, the collector might leave behind information on
which cells are trailed, and which cells contain sharable
information. This would speed up the sharer and in particular the
build phase.

\paragraph{\bf What if Representation Sharing does not work.}

The benchmark programs show that representation sharing is not always
effective: it depends indeed highly on the type of program. When
representation sharing does not work, this can be noticed during a run
of the representation sharing module by observing the hashed\_terms. If it
keeps growing, it means that lots of different terms are found. This
in turn gives an indication that representation sharing is not
effective. An important advantage of our implementation is that the
representation sharing process can be abandoned at any time since no
changes to the WAM run-time data structures are made until the absorb
phase in which structure (or list ...) pointers are
updated, and even the absorb phase can be stopped before
finishing. Also, if representation sharing is run from time to time
only ---as suggested by Ulrich Neumerkel--- then the frequency of
running it can take into account the effectiveness of representation
sharing up to that moment. Such tuning could depend also on the
relative performance of the garbage collector and the representation
sharing module.

\paragraph{\bf Parallelization.}

During the scanning phase, the stacks (heap, local stack ...) are
read-only, while the cached\_hash and the hashed\_terms can be read
and written by different workers.
During the absorb phase, the cached\_hash and
hashed\_terms are read-only, and only the stacks are written to.

By giving different workers a different part of the heap to start
working on, duplicate work might be avoided and synchronization
slowdown kept low in the scanning phase. During the absorb phase,
giving different workers different parts of the stacks makes their
actions completely independent.

\paragraph{\bf Variable Chains.}

\begin{figure}[h]
\begin{centering}
\subfigure[Just before absorbing]{{\epsfig{file=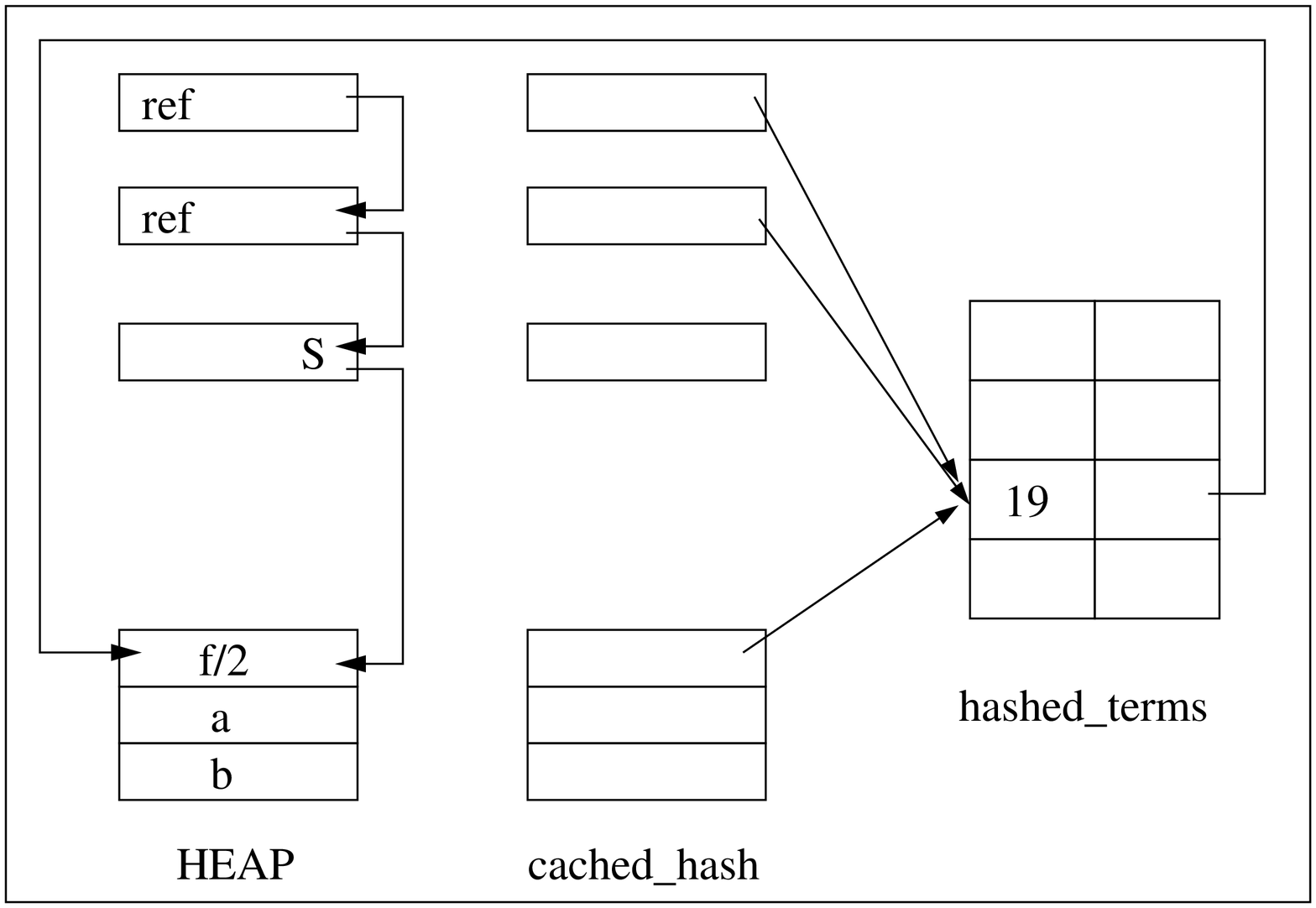,height=0.2\textheight}} \label{chain1}}
\subfigure[After absorbing]{{\epsfig{file=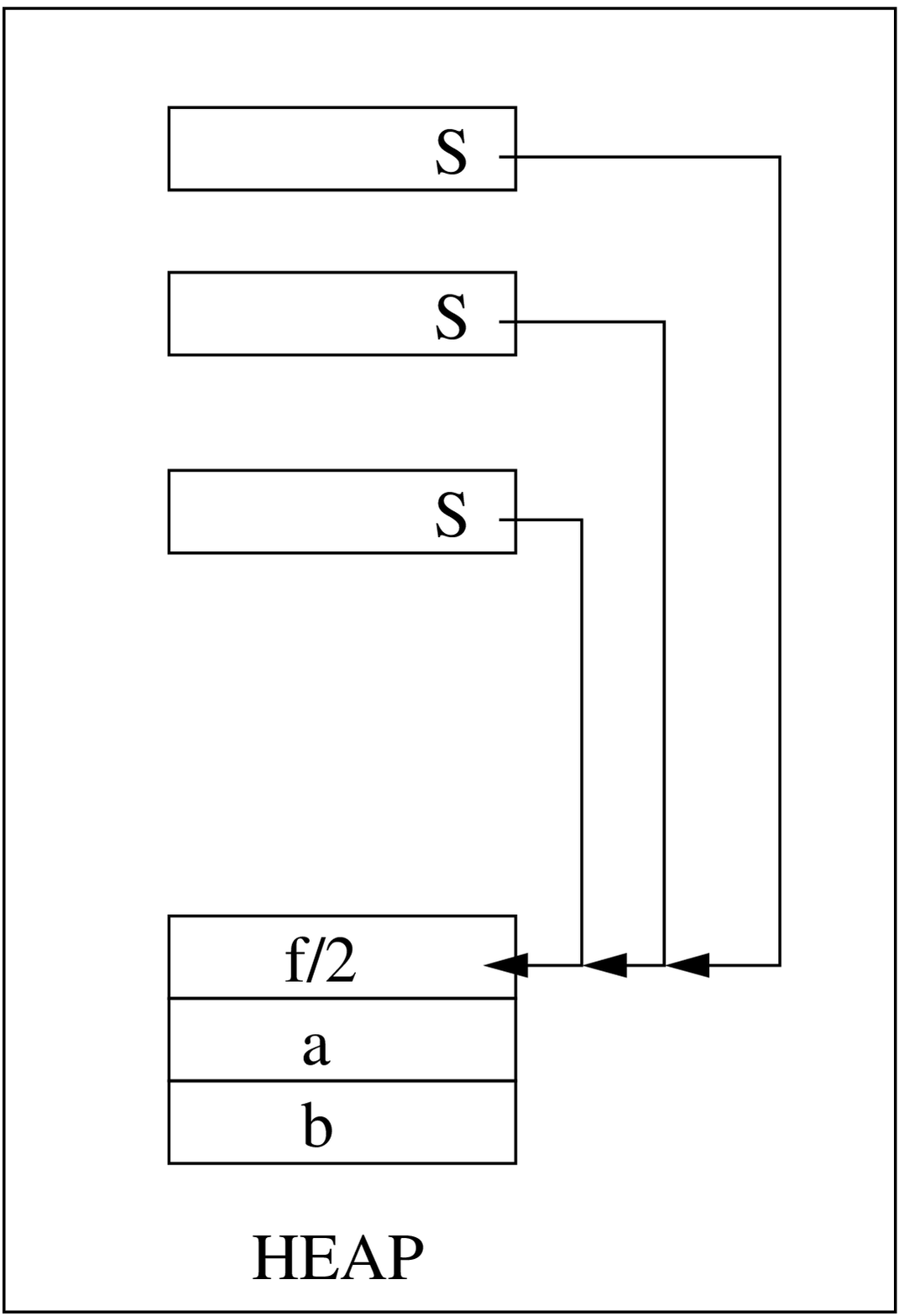,height=0.2\textheight}} \label{chain2}}
\caption{Absorption for chains of references in action}
\label{fig:chain}
\end{centering}
\end{figure}

We have not treated variable chains in much detail, as we were mostly
interested in sharing between the bodies of compound data. However, a
slight extension of the code for the build phase can also call
save\_hash for all reference cells. That results in a similar effect
as variable shunting as described in \cite{VariableShunting}, but is
not as {\em complete} as the method described there. Figure
\ref{fig:chain} shows an example of how a chain of references is
transformed.

\paragraph{\bf Backtrackable Representation Sharing.}\label{backtrackablerepshar}

Backtrackable representation sharing would follow the principle that when
two terms are identical (as for {\em ==/2}) then one can absorb the other,
regardless of whether they have trailed cells or not.  The change made
(to a LIST or STRUCT-tagged pointer) by the absorb phase is now
conditionally (and value) trailed. This costs extra trail space of
course. On cut, the trail can be tidied, so in case the computation
becomes eventually deterministic, the amount of sharing can be
arbitrarily larger than without this form of backtrackable
representation sharing. However, suppose that all sharing were
trailed, then it is possible that an immediately following GC would
not be able to recover anything. And if the computation becomes
deterministic eventually, running the sharer will do the same job as
was done in the case of the backtrackable representation sharing, only
later -- which might be even better, because the earlier sharing could
have been unnecessary because backtracking has destroyed it. All in
all, our feeling is that backtrackable representation sharing is not
worth its while.

\paragraph{\bf Partial Sharing.}

{\em Partial sharing} refers to running the sharer in an incomplete
way, i.e., it achieves part of its potential effect, but maybe not all.

Partial sharing can result for instance from restricting the part of
the heap in which duplicate terms are identified, i.e., restricting the
scan phase to part of the heap. Another possibility is to restrict
sharing to certain terms, e.g., just for lists, or to certain parts
of the other stacks. It is one of the strengths of our implementation
approach that all such variations can be incorporated rather easily.

\paragraph{\bf Incremental Sharing.}

The notion of {\em incremental sharing} refers to the possibility to
perform a partial sharer pass, e.g., on part of the heap, and
continue that pass later on, eventually obtaining the same effect as
running the sharer completely. The ability to perform partial sharing
is certainly needed, but there is more: information must be passed
from one partial run to the other, and the user program and the sharer must
be able to run in an interleaved way. This raises immediately question
of the completeness, but also efficiency is at stake.

The issues with incremental sharing are similar to the ones with
generational sharing in the next paragraph and we do not discuss
further incremental sharing separately.

\paragraph{\bf Generational Sharing.}

The notion of {\em generational sharing} refers to the possibility to
avoid performing sharing on a part of the heap on which it was
performed earlier. In analogy with generational garbage collection,
there is a rationale for performing generational sharing: for
generational garbage collection, the rationale is that new objects
tend to die quickly. For generational sharing the rationale is that
redoing sharing on old data (on which sharing was performed earlier)
does not pay off.

\begin{sloppypar}
Our strategy to non-generational sharing is to recompute the
cached\_hash and hashed\_terms tables from scratch every time after a
new garbage collection. With generational sharing, one would like to
reuse the part of the tables corresponding to the older generation.
\end{sloppypar}

We reason about forward computation first: The information on terms in
the older generation that were ground at the previous run of the
sharer and eligible for sharing at that moment is still valid. The
same is generally not true for a non-ground term: it can now contain
cells that are trailed, and in that case the information about the
term is to be discarded from the table, or at least not used. Since it
is not straightforward to keep track of which information in the tables
is no longer valid because of this reason, it might be best to
restrict a generational sharer to ground terms only.

Now suppose that backtracking has taken place between two activations
of the sharer: generally, this invalidates entries in the sharer
tables because terms have disappeared. It is easy to adapt the
cached\_hash table (it shrinks with the heap on backtracking), but the
hashed\_terms table also needs to be adapted. By keeping high and low
water marks of the top of heap pointer, this can also be achieved.
The cost of adapting the tables might be larger than the cost of
rebuilding them however.

\section{Related Work}\label{related}

\cite{appelhashconsinggc} describes how hash-consing can be performed
during garbage collection in an implementation of Standard ML
(SML/NJ). The collector is generational, and the data structures in
the old generation are hash-consed. In this way, the operation of
hash-consing is restricted to data structures that are expected to
live long. The reported performance and space gains are disappointing:
half of the benchmarks lose performance (up to 25\%) and the gain is
maximally 10\% (for boyer). The space improvement is even smaller: on
most benchmarks less than 1\%. Also for space, boyer is the exception
with about 15\%. Note however, that these space figures are about the
amount of data copied to the older generation, i.e., the data that is
hash-consed, and which is collected infrequently. As such, these
numbers do not give full insight in the potential of hash-consing. Still,
\cite{appelhashconsinggc} is most closely related to our
implementation of representation sharing for Prolog: our strategy is
to perform representation sharing after a (major) garbage collection,
so we introduce sharing only for data that just survived a collection.

Mercury \cite{zoltan:mercury} is basically a functional language, and the
issue of trailing does not enter. In the developers' mailing list in
August 1999, the issue of hash consing was raised with a proposal for an
implementation as well as how to present it to the user. It is
interesting that at some point, the opposite of our {\em :- mutable}
declaration was proposed. As an example {\em :- pragma
hash\_cons(foo/3)} tells the compiler to hash-cons the constructors of
type {\em foo/3}.  As far as we know, the proposals were not implemented.

Last but not least, \cite{neumerkeldiplomarbeit} provides the example
{\em blid/1}, and gives a high-level outline of an algorithm for
minimization of heap terms seen as DFAs. Our implementation can be
seen as a concrete version of that algorithm. However, our {\em
  minimization} shows mostly similarities with \cite{ershov58} in
which Ershov uses (for the first time in the published history of
computer science) hashing to detect common subtrees in a given
tree.

\section{Conclusion}\label{conclusion}

Without the questions by Ulrich Neumerkel on comp.lang.prolog, we
would not have worked on this topic and we are grateful for his
insistence that Prolog systems should have a sharer. We have provided
a practical and efficient implementation of representation sharing,
that can be incorporated without problems in most WAM based
systems. Our implementation has the advantage that it does not rely on
a particular garbage collection strategy or implementation. On the
other hand, a tighter integration of the garbage collector with the
representation sharing module can make the latter more
efficient. Still, representation sharing is not effective for all
programs, so it must not be applied indiscriminately, i.e., it needs
its own policy. We have also shown that input sharing for {\em findall/3} is
easy to implement.

\section*{Acknowledgements}

We thank the anonymous referees: their suggestions have improved the
paper considerably. The second author acknowledges support from
project {\em GOA/08/008 Probabilistic Logic Learning}, and from the Research
Foundation Flanders (FWO) through projects {\em WOG: Declarative
Methods in Computer Science} and {\em G.0221.07 Platform independent
analysis and implementation of Constraint Handling Rules}.



\begin{thebibliography}{}

\bibitem[\protect\citeauthoryear{A{\"\i}t-Kaci}{A{\"\i}t-Kaci}{1991}]{wam:hass%
an}
{\sc A{\"\i}t-Kaci, H.} 1991.
\newblock {\em {Warren's Abstract Machine: A Tutorial Reconstruction}}.
\newblock {MIT} {P}ress.

\bibitem[\protect\citeauthoryear{Appel and Gon{\c c}alves}{Appel and Gon{\c
  c}alves}{1993}]{appelhashconsinggc}
{\sc Appel, A.~W.} {\sc and} {\sc Gon{\c c}alves, M.~J.~R.} 1993.
\newblock {Hash-consing Garbage Collection}.
\newblock Tech. Rep. CS-TR-412-93, Princeton University. Feb.

\bibitem[\protect\citeauthoryear{Appleby, Carlsson, Haridi, and Sahlin}{Appleby
  et~al\mbox{.}}{1988}]{SicstusGarbage@CACM-88}
{\sc Appleby, K.}, {\sc Carlsson, M.}, {\sc Haridi, S.}, {\sc and} {\sc Sahlin,
  D.} 1988.
\newblock {Garbage collection for Prolog based on WAM}.
\newblock {\em Communications of the ACM\/}~{\em 31,\/}~6 (June), 719--741.

\bibitem[\protect\citeauthoryear{Baker}{Baker}{1992}]{bakerwarpspeed}
{\sc Baker, H.~G.} 1992.
\newblock The {B}oyer {B}enchmark at warp {S}peed.
\newblock {\em SIGPLAN Lisp Pointers\/}~{\em V,\/}~3, 13--14.

\bibitem[\protect\citeauthoryear{Bevemyr and Lindgren}{Bevemyr and
  Lindgren}{1994}]{BevemyrLindgren@PLILP-94}
{\sc Bevemyr, J.} {\sc and} {\sc Lindgren, T.} 1994.
\newblock {A simple and efficient copying Garbage Collector for Prolog}.
\newblock In {\em Proceedings of the Sixth International Symposium on
  Programming Language Implementation and Logic Programming}, {M.~Hermenegildo}
  {and} {J.~Penjam}, Eds. Number 844 in Lecture Notes in Computer Science.
  Springer-Verlag, 88--101.

\bibitem[\protect\citeauthoryear{Blockeel, Dehaspe, Demoen, Janssens, Ramon,
  and Vandecasteele}{Blockeel et~al\mbox{.}}{2000}]{querypacks}
{\sc Blockeel, H.}, {\sc Dehaspe, L.}, {\sc Demoen, B.}, {\sc Janssens, G.},
  {\sc Ramon, J.}, {\sc and} {\sc Vandecasteele, H.} 2000.
\newblock {E}xecuting query packs in {ILP}.
\newblock In {\em Inductive Logic Programming, 10th International Conference,
  ILP2000, London, UK, July 2000, Proceedings}, {J.~Cussens} {and} {A.~Frisch},
  Eds. Lecture Notes in Artificial Intelligence, vol. 1866. Springer, 60--77.

\bibitem[\protect\citeauthoryear{Boehm and Weiser}{Boehm and
  Weiser}{1988}]{hansboehm}
{\sc Boehm, H.-J.} {\sc and} {\sc Weiser, M.} 1988.
\newblock Garbage collection in an uncooperative {E}nvironment.
\newblock {\em Software: Practice and Experience\/}~{\em 18}, 807--–820.

\bibitem[\protect\citeauthoryear{Carlsson}{Carlsson}{1990}]{matsphd}
{\sc Carlsson, M.} 1990.
\newblock Design and {I}mplementation of an {O}r-parallel {P}rolog {E}ngine.
\newblock Ph.D. thesis, The Royal Institute of Technology (KTH), Stokholm,
  Sweden. See also: {\sf http://www.sics.se/isl/sicstus.html}.

\bibitem[\protect\citeauthoryear{Chen and Warren}{Chen and
  Warren}{1996}]{ChWa96}
{\sc Chen, W.} {\sc and} {\sc Warren, D.~S.} 1996.
\newblock Tabled {E}valuation with {D}elaying for {G}eneral {L}ogic {P}rograms.
\newblock {\em Journal of the ACM\/}~{\em 43,\/}~1 (January), 20--74.

\bibitem[\protect\citeauthoryear{Clocksin and Mellish}{Clocksin and
  Mellish}{1984}]{ClMe84}
{\sc Clocksin, W.} {\sc and} {\sc Mellish, C.} 1984.
\newblock {\em Programming in Prolog}.
\newblock Springer-Verlag.

\bibitem[\protect\citeauthoryear{Demoen}{Demoen}{2002}]{DemoenICLP2002fresh}
{\sc Demoen, B.} 2002.
\newblock {A} different {L}ook at {G}arbage {C}ollection for the {WAM}.
\newblock In {\em Proceedings of ICLP2002 - International Conference on Logic
  Programming}, {P.~Stuckey}, Ed. Number 2401 in Lecture Notes in Computer
  Science. ALP, Springer-Verlag, Copenhagen, 179--193.

\bibitem[\protect\citeauthoryear{Demoen and Nguyen}{Demoen and
  Nguyen}{2000}]{wamvariations}
{\sc Demoen, B.} {\sc and} {\sc Nguyen, P.-L.} 2000.
\newblock {S}o many {WAM} {V}ariations, so little {T}ime.
\newblock In {\em Computational Logic - CL2000, First International Conference,
  London, UK, July 2000, Proceedings}, {J.~Lloyd}, {V.~Dahl}, {U.~Furbach},
  {M.~Kerber}, {K.-K. Lau}, {C.~Palamidessi}, {L.~M. Pereira}, {Y.~Sagiv},
  {and} {P.~J. Stuckey}, Eds. Lecture Notes in Artificial Intelligence, vol.
  1861. ALP, Springer, 1240--1254.

\bibitem[\protect\citeauthoryear{Ershov}{Ershov}{1958}]{ershov58}
{\sc Ershov, A.~P.} 1958.
\newblock On {P}rogramming of arithmetic {O}perations.
\newblock {\em Commun. ACM\/}~{\em 1,\/}~8, 3--6.

\bibitem[\protect\citeauthoryear{Goto}{Goto}{1974}]{goto74}
{\sc Goto, E.} 1974.
\newblock Monocopy and associative {A}lgorithms in extended {L}isp.
\newblock Tech. Rep. TR 74-03, University of Tokyo.

\bibitem[\protect\citeauthoryear{Grimley-Evans}{Grimley-Evans}{1995}]{findall1%
archive}
{\sc Grimley-Evans, E.} 1995.
\newblock Findall/3 without copying?
\newblock
  http://www.logicprogramming.org/newsletter/archive\_93\_96/net/meta-level/fi%
ndall1.html.

\bibitem[\protect\citeauthoryear{Gutmann and Kersting}{Gutmann and
  Kersting}{2006}]{Gutmann}
{\sc Gutmann, B.} {\sc and} {\sc Kersting, K.} 2006.
\newblock Tildecrf: Conditional random fields for logical sequences.
\newblock In {\em Proceedings of the 15th European Conference on Machine
  Learning (ECML-2006)}, {J.F{\"{u}}rnkranz}, {J.~Scheffer}, {and}
  {T.~Spiliopoulou}, Eds. Springer, 174--185.

\bibitem[\protect\citeauthoryear{Hirsch, Silverman, and Shapiro}{Hirsch
  et~al\mbox{.}}{1987}]{logixFCP}
{\sc Hirsch, M.}, {\sc Silverman, W.}, {\sc and} {\sc Shapiro, E.} 1987.
\newblock Computation {C}ontrol and {P}rotection in the {L}ogix {S}ystem.
\newblock In {\em Concurrent Prolog: Collected Papers, Vols. 1 and 2},
  {E.~Shapiro}, Ed. MIT Press.

\bibitem[\protect\citeauthoryear{Mari{\"{e}}n and Demoen}{Mari{\"{e}}n and
  Demoen}{1993}]{findallwithoutfindall}
{\sc Mari{\"{e}}n, A.} {\sc and} {\sc Demoen, B.} 1993.
\newblock {F}indall without findall/3.
\newblock In {\em Proceedings of the Tenth Int. Conf. on Logic Programming},
  {D.~S. Warren}, Ed. MIT-Press, 408--423.

\bibitem[\protect\citeauthoryear{Neumerkel}{Neumerkel}{1989}]{neumerkeldiploma%
rbeit}
{\sc Neumerkel, U.} 1989.
\newblock Speicherbereinigung f{\"{u}}r {P}rologsysteme.
\newblock M.S.\ thesis, Institut f{\"{u}}r Praktische Informatik der
  Technischen Universit{\"{a}}t Wien.

\bibitem[\protect\citeauthoryear{O'Keefe}{O'Keefe}{2001}]{OKeefePearl}
{\sc O'Keefe, R.} 2001.
\newblock O(1) {R}eversible {T}ree {N}avigation without {C}ycles.
\newblock {\em Theory and Practice of Logic Programming, vol. 1, no. 5, pp 617
  - 630\/}.

\bibitem[\protect\citeauthoryear{Reade}{Reade}{1989}]{ChrisReade}
{\sc Reade, C.} 1989.
\newblock {\em Elements of Functional Programming}.
\newblock Addison-Wesley.

\bibitem[\protect\citeauthoryear{Sagonas, Swift, and Warren}{Sagonas
  et~al\mbox{.}}{1994}]{SaSW94}
{\sc Sagonas, K.}, {\sc Swift, T.}, {\sc and} {\sc Warren, D.~S.} 1994.
\newblock {XSB as an Efficient Deductive Database Engine}.
\newblock In {\em Proc. of SIGMOD 1994 Conference}. Vol. 23:2. ACM, 442--453.

\bibitem[\protect\citeauthoryear{Sahlin and Carlsson}{Sahlin and
  Carlsson}{1991}]{VariableShunting}
{\sc Sahlin, D.} {\sc and} {\sc Carlsson, M.} 1991.
\newblock {{Variable Shunting for the WAM}}.
\newblock Tech. Rep. {SICS/R-91/9107}, {SICS}.

\bibitem[\protect\citeauthoryear{Somogyi, Henderson, and Conway}{Somogyi
  et~al\mbox{.}}{1996}]{zoltan:mercury}
{\sc Somogyi, Z.}, {\sc Henderson, F.}, {\sc and} {\sc Conway, T.} 1996.
\newblock The {E}xecution {A}lgorithm of {M}ercury, an efficient purely
  declarative {L}ogic {P}rogramming {L}anguage.
\newblock {\em Journal of Logic Programming\/}~{\em 29,\/}~1-3, 17--64.

\bibitem[\protect\citeauthoryear{Tarau}{Tarau}{1991}]{Tarau91:JAP}
{\sc Tarau, P.} 1991.
\newblock A {S}implified {A}bstract {M}achine for the {E}xecution of {B}inary
  {M}etaprograms.
\newblock In {\em Proceedings of the Logic Programming Conference'91}. ICOT,
  Tokyo, 119--128.

\bibitem[\protect\citeauthoryear{Tarau}{Tarau}{1992}]{ecologicalPaul@IWMM-92}
{\sc Tarau, P.} 1992.
\newblock {Ecological Memory Management in a Continuation Passing Prolog
  Engine}.
\newblock In {\em Proceedings of IWMM'92: International Workshop on Memory
  Management}, {Y.~Bekkers} {and} {J.~Cohen}, Eds. Number 637 in Lecture Notes
  in Computer Science. Springer-Verlag, 344--356.

\bibitem[\protect\citeauthoryear{Tarau and Majumdar}{Tarau and
  Majumdar}{2009}]{padl09inter}
{\sc Tarau, P.} {\sc and} {\sc Majumdar, A.} 2009.
\newblock {Interoperating Logic Engines}.
\newblock In {\em {Practical Aspects of Declarative Languages, 11th
  International Symposium, PADL 2009}}. Springer, LNCS 5418, Savannah, Georgia,
  137--151.

\bibitem[\protect\citeauthoryear{{Van Roy} and Despain}{{Van Roy} and
  Despain}{1992}]{Aquarius}
{\sc {Van Roy}, P.} {\sc and} {\sc Despain, A.} 1992.
\newblock High-performance {L}ogic {P}rogramming with the {A}quarius {P}rolog
  {C}ompiler.
\newblock {\em IEEE Computer\/}~{\em 25(1)}, 54--68.

\bibitem[\protect\citeauthoryear{Vandeginste, Sagonas, and Demoen}{Vandeginste
  et~al\mbox{.}}{2002}]{VandeginsteSagonasDemoenPADL2002}
{\sc Vandeginste, R.}, {\sc Sagonas, K.}, {\sc and} {\sc Demoen, B.} 2002.
\newblock {S}egment {O}rder preserving and generational {G}arbage {C}ollection
  for {P}rolog.
\newblock In {\em Practical Aspects of Declarative Languages, 4th International
  Symposium, PADL 2002, Proceedings}, {S.~Krishnamurthi} {and}
  {C.~Ramakrishnan}, Eds. Lecture Notes in Computer Science, vol. 2257.
  Springer, 299--317.

\bibitem[\protect\citeauthoryear{Wallace, Novello, and Schimpf}{Wallace
  et~al\mbox{.}}{1997}]{Wallace97eclipse}
{\sc Wallace, M.}, {\sc Novello, S.}, {\sc and} {\sc Schimpf, J.} 1997.
\newblock {ECLiPSe}: A {P}latform for {C}onstraint {L}ogic {P}rogramming.
\newblock {\em ICL Systems Journal\/}~{\em 12,\/}~1 (May), 159--­200.

\bibitem[\protect\citeauthoryear{Warren}{Warren}{1983}]{DHWa83}
{\sc Warren, D. H.~D.} 1983.
\newblock {An Abstract Prolog Instruction Set}.
\newblock Tech. Rep. 309, SRI.

\bibitem[\protect\citeauthoryear{Wielemaker, Huang, and {van der
  Meij}}{Wielemaker et~al\mbox{.}}{2008}]{swiprolog}
{\sc Wielemaker, J.}, {\sc Huang, Z.}, {\sc and} {\sc {van der Meij}, L.} 2008.
\newblock {SWI}-{P}rolog and the {W}eb.
\newblock {\em Theory and Practice of Logic Programming\/}~{\em 8,\/}~3,
  363--392.

\end{thebibliography}



\section*{Appendix: the relevant Part from the Program in \cite{OKeefePearl}}

\begin{Verbatim}[fontsize=\scriptsize, frame=single,samepage=true]
tree_children(node(_,Children), Children).                          up_down_star(A, D) :-
                                                                            (   A = D
top_pointer(Tree, ptr(Tree,[],[],no_ptr)).                                  ;   up_down_plus(A, D)
                                                                            ).
up_down(P, ptr(T,L,R,A)) :-                                         
        (   var(P) ->                                               up_down_plus(A, D) :-
            A = ptr(_,_,_,_), % not no_ptr, that is.                        (   var(A) ->
            P = A                                                               up_down(X, D),
        ;   A = P,                                                              up_down_star(A, X)
            P = ptr(Tree,_,_,_),                                            ;   up_down(A, X),
            tree_children(Tree, Children),                                      up_down_star(X, D)
            % split Children++[] into reverse(L)++[T]++R                    ).
            split_children(Children, [], L, T, R)                   
        ).                                                          mk_tree(D, node(D,C)) :-
                                                                            (   D > 0 ->
split_children([T|R], L, L, T, R).                                              D1 is D - 1,
split_children([X|S], L0, L, T, R) :-                                           C = [T1,T2,T3,T4],
        split_children(S, [X|L0], L, T, R).                                     mk_tree(D1, T1),
                                                                                mk_tree(D1, T2),
f1(N) :- mk_tree(N,T),                                                          mk_tree(D1, T3),
         top_pointer(T,P),                                                      mk_tree(D1, T4)
         findall(Q, up_down_star(P, Q), _L).                                ;   C = []
                                                                            ).
\end{Verbatim}

\end{document}